\documentclass[%
 aip,
 amsmath,amssymb,
 reprint,%
]{revtex4-1}

\usepackage{graphicx}
\usepackage{dcolumn}
\usepackage{bm}

\usepackage[utf8]{inputenc}
\usepackage[T1]{fontenc}
\usepackage{mathptmx}
\usepackage{etoolbox}
\usepackage{caption}
\usepackage{subcaption}

\makeatletter
\def\@email#1#2{%
 \endgroup
 \patchcmd{\titleblock@produce}
  {\frontmatter@RRAPformat}
  {\frontmatter@RRAPformat{\produce@RRAP{*#1\href{mailto:#2}{#2}}}\frontmatter@RRAPformat}
  {}{}
}%
\makeatother
\begin{document}

\preprint{AIP/123-QED}

\title[Photo-induced spall failure of (111) twist grain boundaries in Ni bicrystals]{Photo-induced spall failure of (111) twist grain boundaries in Ni bicrystals}
\author{M. Isiet}
 \affiliation{Department of Mechanical Engineering, University of British Columbia, Vancouver, Canada}
\author{M. Ponga\textsuperscript{*}}%
 \email{mponga@mech.ubc.ca}
\affiliation{Department of Mechanical Engineering, University of British Columbia, Vancouver, Canada}%

\date{\today}

\begin{abstract}
Spall failure, a complex failure mechanism driven by tensile stress wave interactions, has been extensively studied in single-crystal FCC metals, revealing a precursor stage involving dislocation emission along closed-packed directions. 
Here we investigate the photo-induced spall failure of Ni bicrystals under a two-pulse laser configuration, exploring various misorientation angles through two-temperature molecular dynamics (MD) simulations including electronic effects to simulate light-matter interaction. 
Our findings demonstrate that light-matter interactions can induce spall failure at the sample center, similar to conventional plate-impact methods, when two laser-pulses are applied to the front and back surfaces of the sample. 
The study reveals the significant influence of misorientation angles on dislocation activity and spall behavior, where grain boundaries (GBs) play pivotal roles, either promoting or impeding dislocation interactions. 
Furthermore, our work highlights the potential for enhancing spall resistance by tailoring materials through misorientation angle variation.
\end{abstract}

\maketitle
\textbf{Photo-induced} loading of metallic materials provides a valuable opportunity to investigate the influence of interfaces on strength and failure response. 
This is attributed to the generation of shockwaves with strain rates exceeding $10^8 \mathrm{s}^{-1}$, as demonstrated in previous studies \cite{deRessguier2017, Kanel2010, Echeverria2021,1st_paper}.
The excitation of conduction band electrons and rapid transfer of energy to the phonons result in a rapid temperature increase, leading to the formation of a rapidly expanding plasma. 
Subsequently, compressive waves, followed by unloading tensile waves, are generated at the loaded surface via the rocket effect \cite{Remington2018}. 
These shockwaves propagate away from the irradiated surface and, upon reaching the opposite end, reflect as tensile stress waves to satisfy the traction-free boundary condition. 
Interaction between these tensile waves results in the attainment of a state of hydrostatic pressure. 
If this pressure exceeds the material strength, it triggers a unique dynamic failure mechanism known as spallation, and is driven by ductile failure, i.e. void nucleation, growth, and coalescence \cite{Ariza2011, Ponga2016, Grgoire2017, Bringa2010, Remington2006}.

Understanding the influence of interfaces on spall strength and failure response is crucial for comprehending the complete mechanical behavior of materials, as they naturally exist due to the presence of multiple grains with varying orientations.
Grain boundaries (GBs) are a type of interface that forms between neighbouring grains, i.e., single crystals, in a polycrystalline material.
Each grain is characterized by a grain boundary plane and misorientation angle with respect to each other.
Dynamic and quasi-static experiments on face centered-cubic (fcc) metals have shown that void nucleation is preceded by dislocation emission at grain boundaries \cite{Kumar2003_1}, and that certain grain boundaries are more resistant to failure \cite{Fensin2014}.
Furthermore, it has been shown grain boundary type and not just the grain orientation on either side of the boundary is important to early stage dynamic damage \cite{Cerreta2012}.
This finding has been supported by other works where it was found that grain boundaries with specific misorientations act as preferential sites for defect formation \cite{Wayne2010, Escobedo2011, Escobedo2013,Euser2023}.
However, due to very small temporal (\textbf{fs}--\textbf{ns}) and spatial scales (\textbf{nm}), \textit{in-situ} observation of these failure mechanism has been an experimental challenge.
As a result, molecular dynamics (MD) has been used as a tool to uncover the real-time atomistic insights on the role of interfaces under \textit{plate impact} shock \cite{Fensin2014, Konnur2021, Zhu2022}.
In MD studies, coherent grain boundaries are often constructed according to the theory of coincident site lattice (CSL) \cite{Konnur2021,Long2020,Fensin2014}, where certain misorientation angles produce a periodic array of lattice points in the interface.
Furthermore, to understand the underlying failure mechanism, bicrystals are typical studied by bonding two different grains with each other.
Bicrystalline systems offer precise control over degrees of freedom, facilitating detailed analysis of GB structure, dislocation nucleation, and deformation behavior.

Therefore, in this work, we investigate the laser shock behavior of coherent GBs in Ni bicrystals using a two laser-pulse setup as proposed in our recent work \cite{1st_paper}, to understand the role of GBs in spall failure. 
The two-pulse laser setup has been proposed as an alternative to the plate impact method where the two surfaces of the target are illuminated leading to the generation of a compression front followed by the release of unloading tensile waves at both ends \cite{1st_paper}.
Experiments and atomistic simulations revealed that the interaction of the unloading tensile waves can take place in the middle of the specimen leading to a state of pure hydrostatic tensile stress , and as such, laser-induced spall failure could be achieved at arbitrary locations in the sample \cite{1st_paper}.
Here, the local two-temperature model ($\ell$2T-MD) \cite{Ponga2018, Ullah2019, Hendy2023} was used to incorporate light-matter interactions and resolve electronic effects.
The selection of the (111) GB plane was based on its characteristic as the close-packed plane in fcc metals, which facilitates stacking fault formation and dislocation motion \cite{Long2020}. 
Additionally, the chosen orientation of the GB is perpendicular to the direction of shock propagation, allowing for a comprehensive investigation of the GB response under shock loading.
Within our MD study, we analyze the transition in \textbf{photo-induced damage} and investigate the behavior of CSL configurations under increasing fluence, while also examining the relationship between grain boundary, energy, and spall strength.
\begin{figure*}
\centering
	\begin{subfigure}[b]{0.55\textwidth}
		\centering
		\includegraphics[width=\linewidth]{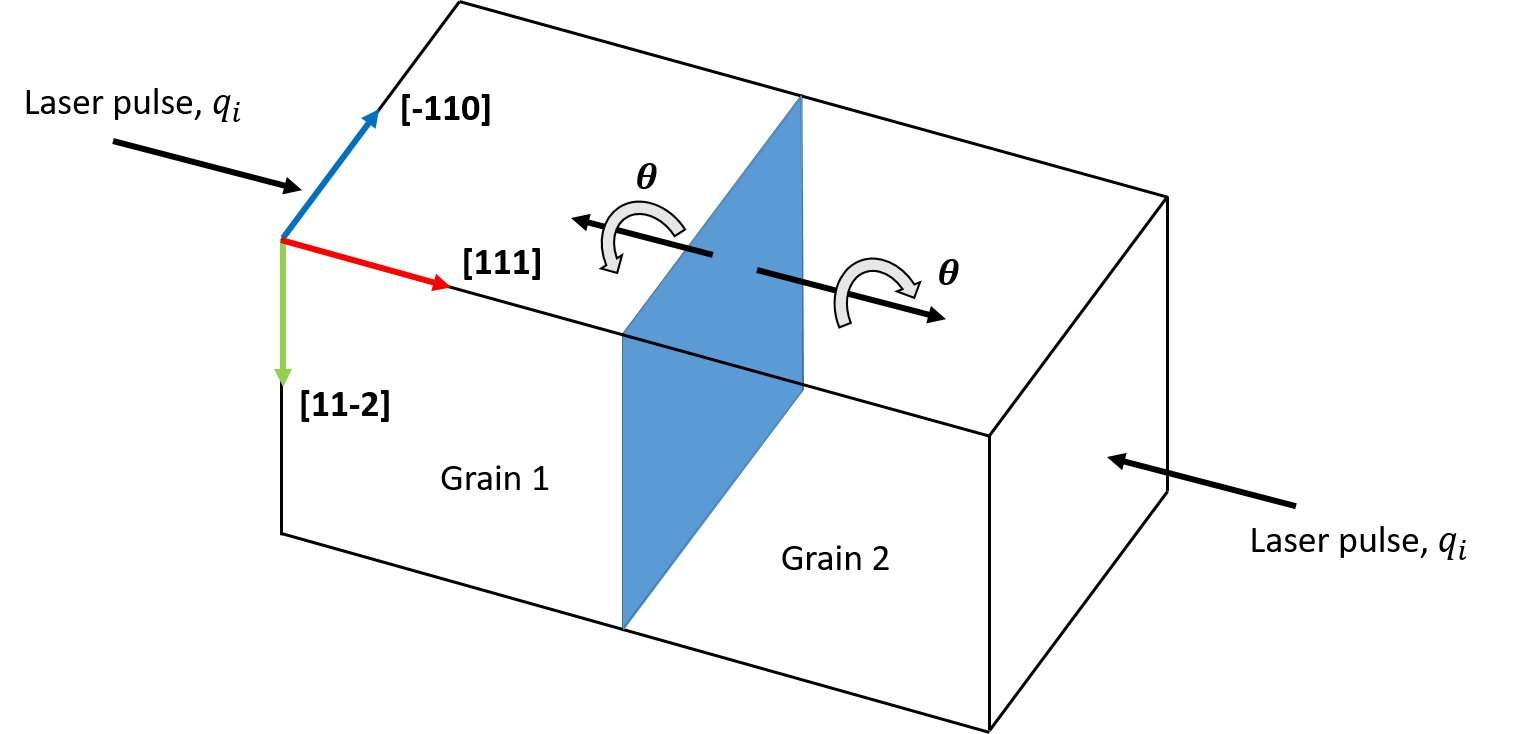}
		\caption{}
	\end{subfigure}
	\begin{subfigure}[b]{0.25\textwidth}
		\centering
		\includegraphics[width=\linewidth]{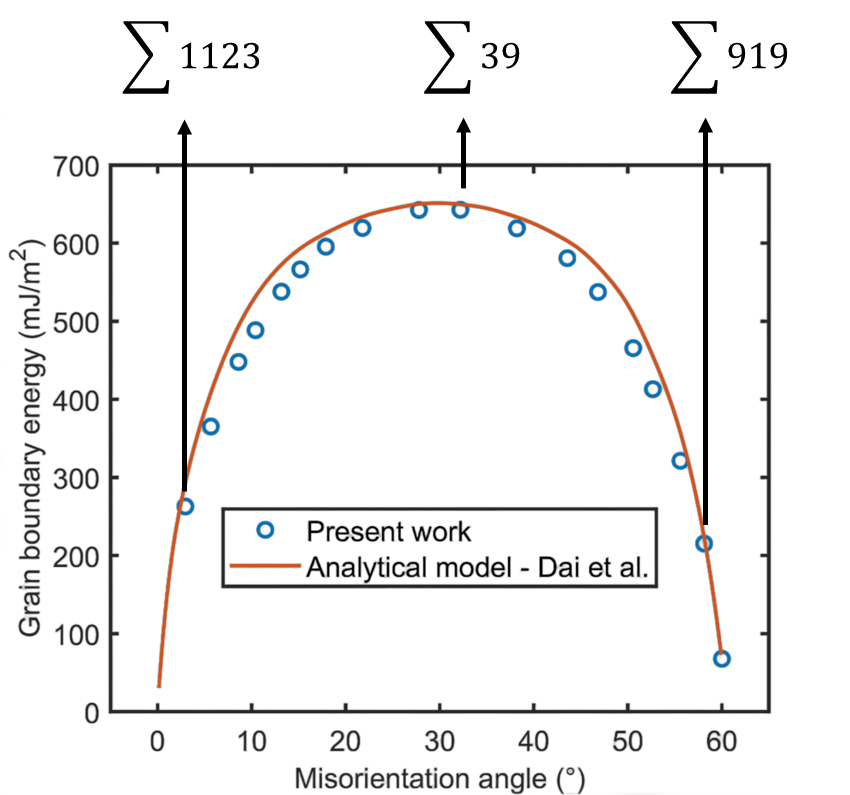}
		\caption{}
	\end{subfigure}
	\begin{subfigure}[b]{0.8\textwidth}
		\centering
		\includegraphics[width=\linewidth]{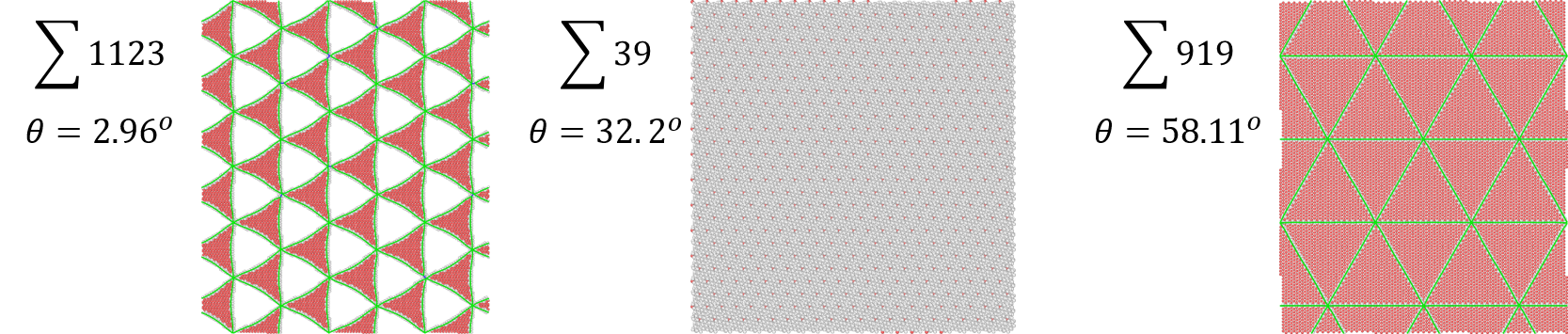}
		\caption{}
	\end{subfigure}    
    \caption{Panel figure illustrating various aspects of [111] twist grain boundaries (GBs) in Ni. (a) MD schematic representation of Ni bicrystal irradiated from both ends with the GB orientated along the direction of the laser pulse. (b) Plot of misorientation angle ($\theta$) with GB energy, highlighting the symmetry around $\theta = 30^\circ$. (c) Atomic structures of [111] twist boundaries in Ni, with each image representing the simulation cell size in the boundary plane. Color coding indicates the Common Neighbour Analysis (CNA) values, with white, red, and green representing defect, HCP, and FCC atoms, respectively.} \label{fig:GB_panel}
\end{figure*}

The spall behavior of Ni bicrystals was investigated using MD simulations using the Large-scale Atomic/Molecular Massively Parallel Simulator (LAMMPS) \cite{Plimpton1995,LAMMPS}.
Ni bicrystals were created with the [111], [11-2], and [-110] crystallographic directions of the face-centered cubic phase aligned with the $x$, $y$, and $z$ axes of the simulation cell, as shown in Figure \ref{fig:GB_panel}(a).
The embedded atom method (EAM) \cite{Daw1984} has been widely employed to model metallic systems, and here, we used a Ni EAM \cite{Mishin2004} that yielded grain boundary energy consistent with calculated and experimental measured energies \cite{Dai2014} as demonstrated in Figure \ref{fig:GB_panel}(b).
Further details on the method used to obtain the 0~K minimum energy grain boundary structures can be found from the literature \cite{Tschopp2007_1, Tschopp2007_2}.
Since the rotating axis of the (111) twist GB lies in the [111] axis, a three-fold symmetry exists, leading to a reduction in the potential misorientation angles ($\theta$) to 0\textdegree -- 120\textdegree.
Additionally, as the twist GBs remain identical when rotated clockwise and counter-clockwise, the potential $\theta$ values are further limited to 0\textdegree -- 60\textdegree.
Within this range, three types of GBs can be obtained \cite{Dai2014,Long2020}: (a) low-angle GBs (LAGB, 0\textdegree -- 12\textdegree), (b) intermediate-angle GBs (IAGB, 12\textdegree -- 48\textdegree), and (c) near-twin GBs (NTGB, 48\textdegree -- 60\textdegree).
These configurations were constructed using the python-based GB code \cite{Hadian2018}.
The generated simulation cells contained between 6.8 to 13.2 $\times 10^{6}$ with a length of 200 nm, while the dimensions in the $y$ and $z$ directions varied between 20 to 30 nm depending on the CSL value.
The generated GB is located at the centre of the simulation cell at the interface between the two equi-sized grains.

A traction-free boundary condition was applied in the direction parallel to the laser pulse ($x$) and periodic boundary conditions were applied to the transverse ($y$ and $z$) directions.
To achieve a comprehensive relaxation of GB structure, a computational annealing process based on an isothermal-isobaric ensemble (NPT) \cite{Martyna1994} was employed.
The annealing procedure \cite{Hasnaoui2004, Hendy2023} was designed to achieve equilibrium and stability of the GB. 
It involved a series of steps to thoroughly relax the GB, spanning from 300 K to 1000 K, over a total duration of 500 ps.
The material constants used for describing the electron subsystem in the $\ell$2T-MD model were obtained from the literature \cite{1st_paper, Hendy2023}.

Figure \ref{fig:GB_panel}(c) presents the observed dislocation profile of the bicrystals.
In the case of a LAGB ($\Sigma 1123$), a set of partial dislocations intersects at triple-junction nodes leading to the formation of triangular networks at the GB.
Both LAGB and NTGB ($\Sigma 919$) exhibit similar dislocation profiles, however subtle differences exist.
Along LAGBs, partial dislocations separate \textit{extrinsic} stacking fault regions from the perfect FCC crystal regions.
Conversely, NTGBs feature partial dislocations separating \textit{intrinsic} stacking faults.
As for IAGB ($\Sigma 39$), the dislocation structure is not readily discernible due to convergence of dislocation cores causing dislocation lines to impinge and annihilate.
These findings regarding GB structures are in agreement with previous simulation studies \cite{Dai2014, Long2020, Feng2015}.

The relaxed bicrystal samples were then subjected to the two-pulse laser shock at a time-step of 0.1 ps to ensure numerical stability.
A set of absorbed laser fluence, $F_{\mathrm{abs}} = F(1-R)$, including 1000, 1250, 1500, 1750, and 2000 $\mathrm{J}\cdot \mathrm{m^{-2}}$ with pulse duration, $t_{p}$, of 0.1 ps and an initial temperature of 300 K was used.
Local atoms deformation and structures were assessed using the common neighbour analysis (CNA) \cite{Honeycutt1987} and dislocation extraction algorithm (DXA) \cite{Stukowski2010}, both employed with OVITO \cite{Stukowski2009}.
Physical quantities, such as velocity, temperature, density, and stress fields, were determined via a one-dimensional binning analysis along the $x$-direction by averaging the atoms within each bin over brief time intervals \cite{Rojas2021}.
\begin{table*}
\caption{\label{tab:CSL_spall_strength}Relationship between different (111) twist Ni GBs with initial dislocation density ($\rho_d$), fluence threshold for spallation ($F_{\mathrm{abs}}^{\mathrm{spall}}$), shock pressure ($P^{\mathrm{spall}}_{\mathrm{shock}}$), and spall strength ($\sigma^{\mathrm{spall}}$).}
\begin{ruledtabular}
\begin{tabular}{lcccccc}
\multicolumn{1}{c}{GB Type} &\multicolumn{1}{c}{$\theta (^\circ)$} &\multicolumn{1}{c}{$\Sigma$} &\multicolumn{1}{c}{$\rho_d (\times \mathrm{10^{14}}$ $\mathrm{m}^{-2})$}&\multicolumn{1}{c}{$F_{\mathrm{abs}}^{\mathrm{spall}}$ ($\mathrm{J}\cdot \mathrm{m^{-2}}$)} &\multicolumn{1}{c}{$P^{\mathrm{spall}}_{\mathrm{shock}}$ (GPa)}&\multicolumn{1}{c}{$\sigma^{\mathrm{spall}}$ (GPa)}\\ \hline \hline
No GB & 0.0 & 1 & 0 & 1750 & 10.89 & 11.37 \\
LAGB & 2.96 & 1123 & 4.10 & 1750 & 8.68 & 12.86 \\
     & 5.67 & 307 & 2.48 & 1750 & 9.01 & 13.42 \\
     & 8.61 & 133 & 4.10 & 1750 & 12.88 & 13.61 \\
     & 10.42 & 91 & 4.85 & 1750 & 12.38 & 13.81 \\ 
IAGB & 13.17 & 57 & 6.27 & 1750 & 12.83 & 14.00 \\
     & 15.18 & 43 & 6.84 & 1750 & 9.95 & 13.96 \\
     & 17.90 & 31 & 7.18 & 1750 & 12.19  & 14.18 \\
     & 21.79 & 21 & 8.28 & 1500 & 11.88 & 13.78 \\
     & 27.80 & 13 & 0 & 1500 & 7.09 & 13.74 \\
     & 32.20 & 39 & 0 & 1500 & 7.32 & 13.40 \\
     & 38.21 & 7 & 0.03 & 1500 & 10.53 & 12.98 \\
     & 43.57 & 49 & 7.24 & 1750 & 12.35 & 13.89 \\
     & 46.83 & 19 & 6.24 & 1500 & 8.15 & 13.84 \\
NTGB & 50.57 & 37 & 4.61 & 1750 & 12.60 & 13.46 \\ 
     & 52.66 & 61 & 3.44 & 1750 & 12.24 & 13.25 \\ 
     & 55.59 & 169 & 2.48 & 1750 & 8.63 & 12.93 \\ 
     & 58.11 & 919 & 8.70 & 1750 & 9.90 & 12.54 \\ 
     & 60.00 & 3 & 0.03 & 1750 & 12.88 & 11.79 \\
\end{tabular}
\end{ruledtabular}
\end{table*}
\begin{figure*}
\centering
	\begin{subfigure}[b]{0.4\textwidth}
		\centering
		\includegraphics[width=\linewidth]{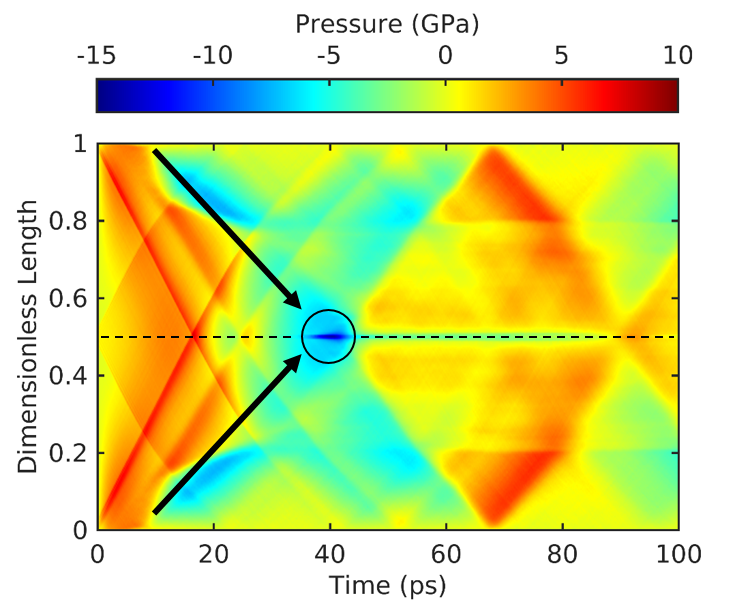}
		\caption{}
	\end{subfigure}
	\begin{subfigure}[b]{0.4\textwidth}
		\centering
		\includegraphics[width=\linewidth]{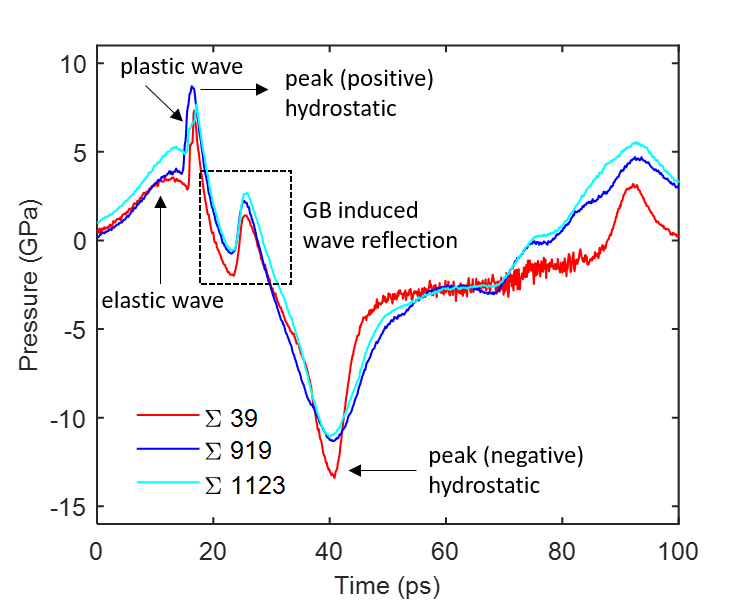}
		\caption{}
	\end{subfigure}
	\begin{subfigure}[b]{0.8\textwidth}
		\centering
		\includegraphics[width=\linewidth]{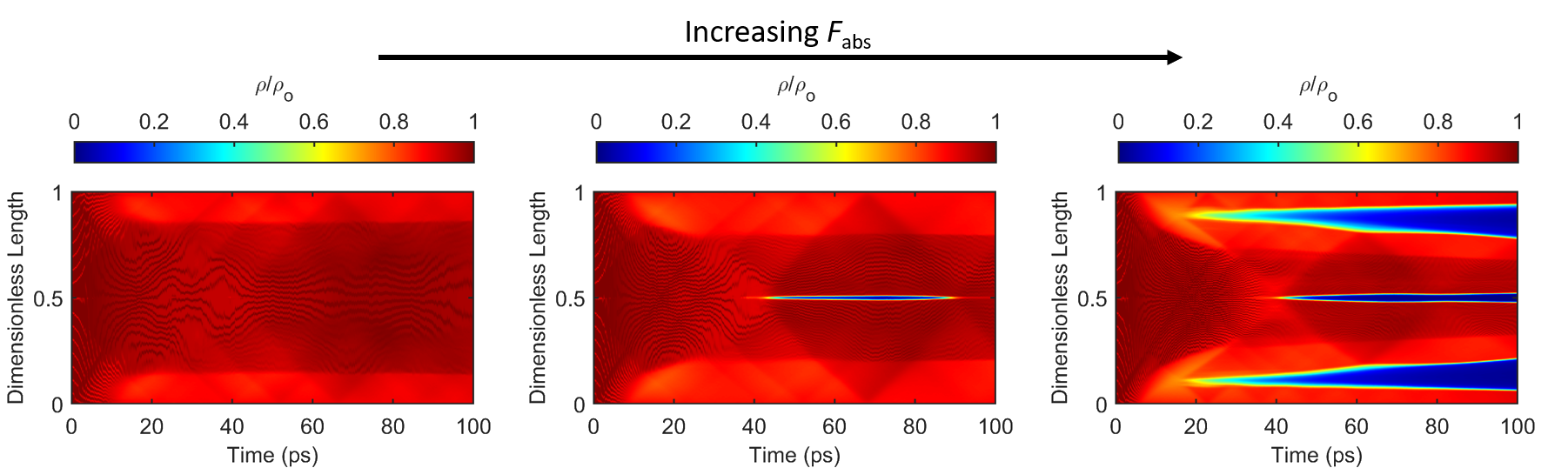}
		\caption{}
	\end{subfigure}
    \caption{Schematic of the light-matter interaction in (111) twist Ni bicrystals, showing (a) the pressure profile as a function of space and time for a IAGB ($\Sigma 39$) under a laser fluence of 1500 $\mathrm{J}\cdot \mathrm{m^{-2}}$, (b) the pressure profile at the grain boundary for a LAGB, IAGB, and NTGB, and (c) the density profile to demonstrate the transition in material behaviour with varying laser fluence, 1250 $\mathrm{J}\cdot \mathrm{m^{-2}}$ of 1500 $\mathrm{J}\cdot \mathrm{m^{-2}}$, and 2000 $\mathrm{J}\cdot \mathrm{m^{-2}}$ for a IAGB ($\Sigma 39$).} \label{fig:shock_behaviour}
\end{figure*}

Pressure evolution induced by the interaction of the two laser pulses, depicted in Figure \ref{fig:shock_behaviour}(a), demonstrates the sequence of compression, tension, and spall failure phenomena experienced in a IAGB ($\Sigma 39$) under a laser fluence of 1500 $\mathrm{J}\cdot \mathrm{m^{-2}}$.
The color gradient, with red and blue regions, represents significant compressive and tensile stresses, respectively. 
The rapid deposition initiates compressive pressure buildup as a result of stress confinement, which, in turn, relaxes by driving the compressive waves away towards the centre of the sample where the GB is located (indicated by the dashed line in Figure \ref{fig:shock_behaviour}(a)).
As two pulses are utilized, symmetrical effects take place at both ends.
Additionally, the release of unloading tensile waves occurs as the pressure relaxes at the loaded surfaces.
These waves propagate towards each other and upon superimposition, a state of large hydrostatic tensile stress is achieved $\sim 15$ GPa, which develops void nucleation as shown in the solid circle in Figure \ref{fig:shock_behaviour}(a).
The behavior of the unloading waves has been thoroughly discussed and analytically calculated in our prior work \cite{1st_paper}.
The pressure profile along the direction of the laser is displayed in Figure \ref{fig:shock_behaviour}(b) where maximum compressive ($\sim$20 ps) and tensile stress was reached ($\sim$40 ps).
Here, we can see that changing $\theta$, i.e., from LAGB ($\Sigma 1123$, $\theta = 2.96^\circ$) to IAGB ($\Sigma 39$, $\theta = 32.2^\circ$) to NTGB ($\Sigma 919$, $\theta = 58.11^\circ$), impacts the peak hydrostatic stress. 
Furthermore, we can see GB induced wave reflection due to the propagation of the elastic and plastic shockwaves followed by the interaction of the unloading tensile waves at the GB. 
Hence, we can see that degree of misorientation may have an impact on the material response as the interaction of the compressive and tensile waves at the GB affects the emission of dislocation and defects which is a precusor to spallation \cite{Galitskiy2018, Pang2014}.
We also observe transition in photo-induced damage with varying laser fluence as demonstrated in Figure \ref{fig:shock_behaviour}(c).
Here, we have the time-averaged density profile of the sample, where the color gradient, red to blue, indicates whether the sample is fully solid (when $\rho/\rho_0 \rightarrow 1$) and the nucleation of voids (when $\rho/\rho_0 \rightarrow 0$).
It is observed that a transition from no failure ($F_{\mathrm{abs}} =$ 1250 $\mathrm{J}\cdot \mathrm{m^{-2}}$) to spallation ($F_{\mathrm{abs}} =$ 1500 $\mathrm{J}\cdot \mathrm{m^{-2}}$) to a mixture of ablation and spallation ($F_{\mathrm{abs}} =$ 2000 $\mathrm{J}\cdot \mathrm{m^{-2}}$) takes place with increasing laser fluence \cite{Shugaev2021}.
When the laser interacts with the surfaces of the sample, a region corresponding to the absorption depth tends to absorb the laser energy leading to an increase in the electronic temperature \cite{Li2020}.
Once this energy is equilibrated by the electrons, within femtoseconds, the energy is transferred to the lattice resulting in an increase in both temperature and pressure.
Hence, we see here that upon exceeding the fluence threshold for spallation, the unloading tensile waves nucleate voids close to the irradiated surfaces due to the reduced stress-bearing capabilities of the metal \cite{de2010dynamic,1st_paper}.
Void nucleation at both the solid and molten regions occurs due to sufficient pressure buildup leading to spallation at the centre of the sample where the unloading tensile wave superimpose.
This transition in light-matter behaviour was similarly seen for all the considered GBs.

As it was seen that no spall damage occurred when $F_{\mathrm{abs}} < 1500$ $\mathrm{J}\cdot \mathrm{m^{-2}}$ as displayed in Table \ref{tab:CSL_spall_strength}, we shift our focus to 1500 and 1750 $\mathrm{J}\cdot \mathrm{m^{-2}}$.
Due to strong photo-induced shocks, GB plasticity is triggered leading to defect formation and dislocation emission at the GB.
Distinct $\theta$ values for twist GBs indicate varying fluence threshold for spallation, implying a variation in spall resistance.
As shown in Figure \ref{fig:shock_behaviour}(a), when $F_{\mathrm{abs}} = F_{\mathrm{abs}}^{\mathrm{spall}}$, a critical tensile stress is reached at the spall plane inducing void nucleation.
Moreover, it is noteworthy that $\sigma^{\mathrm{spall}}$ is generally largest in IAGBs compared to LAGBs and NTGBs.
It is observed here that a $\sim 25 \%$ improvement in spall strength was achieved when comparing all the considered GB structure, highlighting the importance of misorientation angle with spall resistance. 
Additionally, GBs in Ni demonstrate greater spall failure resistance compared to the bulk, as evident from the higher $\sigma^{\mathrm{spall}}$ values of Ni bi-crystals relative to the single-crystal reference structure, $\Sigma 1$ (see Table \ref{tab:CSL_spall_strength}).
We see here that, opposed to conventional notion, lower GB energy does not correlate to stronger GBs, suggesting that GB energy might not be the primary factor influencing the spall behaviour of twist GBs.
These outcomes align with plate impact MD simulations on twist Cu GBs \cite{Fensin2014, Long2020}, underlining the significance of the shock response of the GB and role of initial dislocation network in shaping overall spall behavior.
\begin{figure*}
\centering
	\begin{subfigure}[b]{0.5\textwidth}
		\centering
		\includegraphics[width=\linewidth]{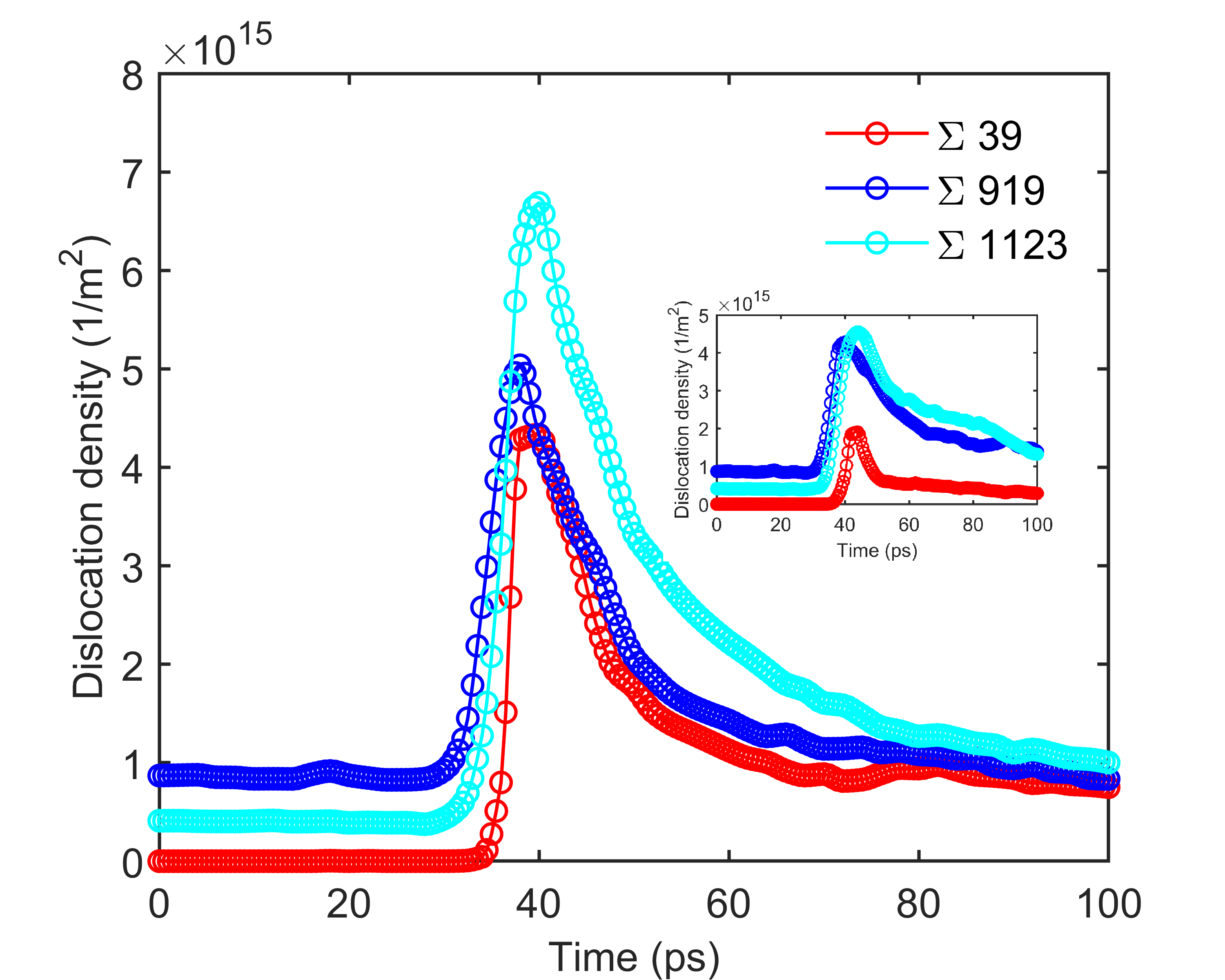}
		\caption{}
	\end{subfigure}
	\begin{subfigure}[b]{0.3\textwidth}
		\centering
		\includegraphics[width=\linewidth]{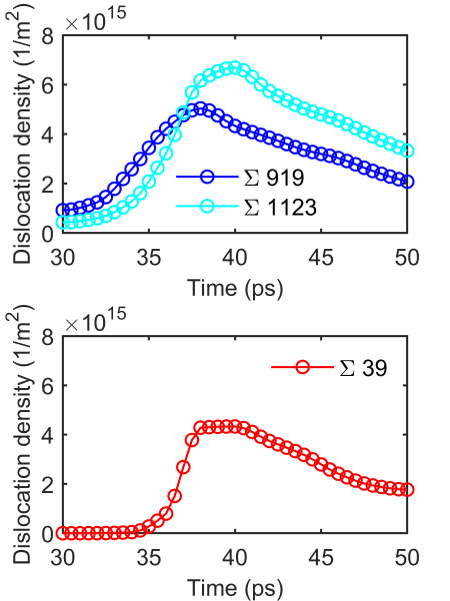}
		\caption{}
	\end{subfigure}
    \caption{Dislocation activity with (a) temporal profile of dislocation density for a LAGB ($\Sigma 1123$, $\theta = 2.96^\circ$), IAGB ($\Sigma 39$, $\theta = 32.2^\circ$), and NTGB ($\Sigma 919$, $\theta = 58.11^\circ$) at 1750 $\mathrm{J}\cdot \mathrm{m^{-2}}$ (inset figure is for the case of 1500 $\mathrm{J}\cdot \mathrm{m^{-2}}$) and (b) focuses on the temporal period where the tensile waves meet at the GB. Here the LAGB and NTGB have been grouped together due to similar trends in comparison to the IAGB.} \label{fig:dislocation_temporal_profile}
\end{figure*}

Next, we examine dislocation behavior at three GBs: LAGB ($\Sigma 1123$, $\theta = 2.96^\circ$), IAGB ($\Sigma 39$, $\theta = 32.2^\circ$), and NTGB ($\Sigma 919$, $\theta = 58.11^\circ$) at their respective $F_{\mathrm{abs}}^{\mathrm{spall}}$. 
The temporal profile of dislocation density in Figure \ref{fig:dislocation_temporal_profile}(a) reveals a consistent pattern across these GBs.
During the compression stage, dislocation density remains nearly constant, with dislocations either growing or annihilating at the GB.
Subsequently, a peak dislocation density corresponding to the maximum tensile stress occurs around $\sim$40 ps.
Notably, the peak values differ among the three cases, with LAGB exhibiting the highest dislocation density, followed by NTGB, and then IAGB.
Interestingly, despite having a lower peak dislocation density, IAGB experiences spallation at a lower $F_{\mathrm{abs}}^{\mathrm{spall}}$ for most cases, indicating a complex interaction between laser pulse, electrons, and phonons with respect to the lattice orientation. 
Once the peak value is reached, a sharp reduction in dislocation density coincides with spall failure onset and the propagation of tensile waves away from the spall region.
The sharp drop aligns with previous observations in imperfect Ni samples with initial voids compared to perfect Ni samples \cite{1st_paper}.
Imperfections are known to serve as prime sites for failure \cite{Remington2018}.
Furthermore, as discussed earlier, $F_{\mathrm{abs}}$ influences the type of laser-induced damage and controls pressure buildup, directly impacting dislocation emission (cf., Figure \ref{fig:dislocation_temporal_profile} with inset figure). 
This highlights that GB type also plays a crucial role in spallation susceptibility.
Therefore, analyzing dislocation activity at the GB is vital for comprehending the dynamics of these interfaces.
\begin{figure*}
\centering
	\begin{subfigure}[b]{0.8\textwidth}
		\centering
		\includegraphics[width=\linewidth]{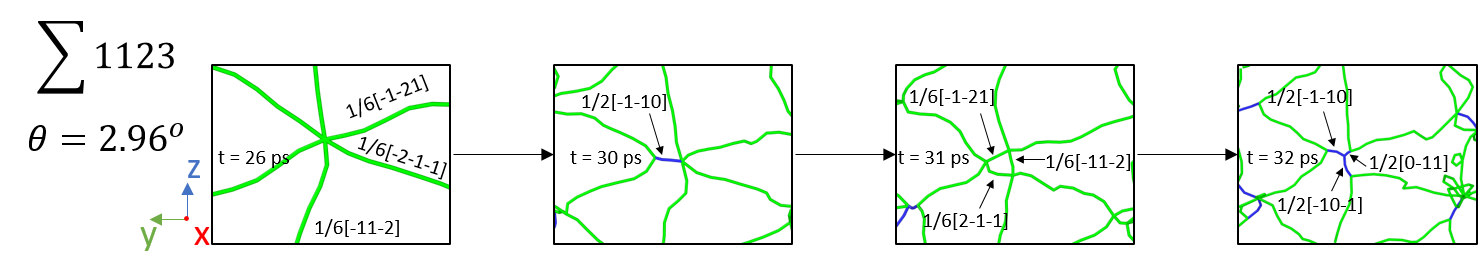}
		\caption{}
	\end{subfigure}
	\begin{subfigure}[b]{0.8\textwidth}
		\centering
		\includegraphics[width=\linewidth]{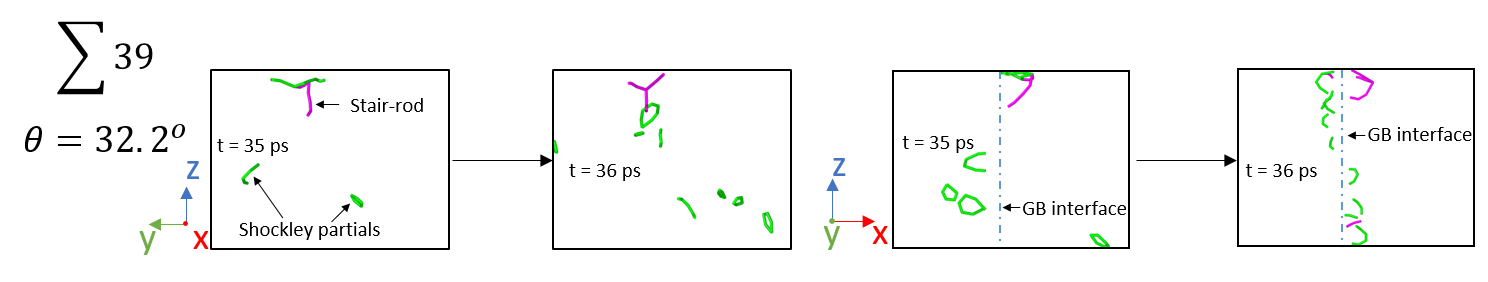}
		\caption{}
	\end{subfigure}
	\begin{subfigure}[b]{0.8\textwidth}
		\centering
		\includegraphics[width=\linewidth]{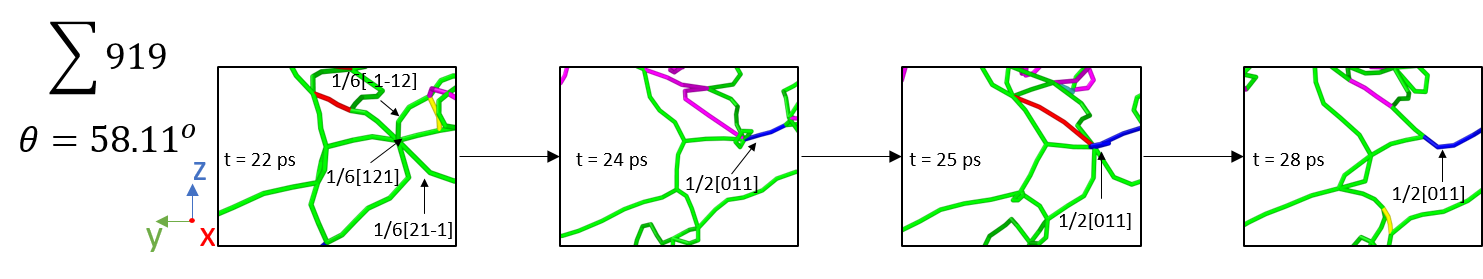}
		\caption{}
	\end{subfigure}
    \caption{Schematic of the evolution of the initial dislocation for (a) LAGB ($\Sigma 1123$, $\theta = 2.96^\circ$), (b) IAGB ($\Sigma 39$, $\theta = 32.2^\circ$), and (c) NTGB ($\Sigma 919$, $\theta = 58.11^\circ$) at their respective $F_{\mathrm{abs}}^{\mathrm{spall}}$ where green, blue, magenta, yellow, and red represents Shockley partials, perfect, stair-rod, Hirth, and other dislocations, respectively.} \label{fig:dislocation_behaviour}
\end{figure*}

Figure \ref{fig:dislocation_behaviour} illustrates the evolution of the initial dislocation over time for the three types of GB.
In the case of the LAGB (Figure \ref{fig:dislocation_behaviour} (a)), GB plasticity initiates at the triple-junction nodes.
Shock compression leads to the formation of perfect dislocations, which annihilate partial dislocations to reduce the strain energy.
Subsequently, more Shockley partials form at the triple junctions as atoms adjust to energetically favorable positions. 
This process continues until the tensile waves reach the GB.
For IAGB, dislocation density increases around $\sim$35 ps, coinciding with the tension phase at the GB (see Figure \ref{fig:shock_behaviour}(a)). 
Shockley partials are emitted at the GB, but the IAGB's initial structure, characterized by a layer of defect atoms (Figure \ref{fig:GB_panel}(c)), acts as a barrier between the grains. 
Consequently, Shockley partials tend to interact with each other at their respective grains. 
Despite its higher GB energy, the IAGB's initial GB structure has little influence during the compression stage.
However, the elevated GB energy reduces the energy threshold for plasticity nucleation \cite{Kumar2003, Long2020}, enabling dislocation emission without pre-existing dislocations.
As can be seen in Figure \ref{fig:dislocation_behaviour}(c), the onset of GB plasticity in NTGB occurs at the intersection between the intrinsic stacking faults similar to LAGB. 
Here also, the triple junction nodes serves as a seed for the formation of perfect dislocations, however instead of emitting Shockley partials, perfect dislocations tend to grow and expand.
However due to possessing fewer triple junction nodes in comparison (see Figure \ref{fig:GB_panel}(c)), dislocation activity in the NTGB is lower than the LAGB resulting in a lower peak dislocation density despite having a larger initial dislocation density (see Figure \ref{fig:dislocation_behaviour}(a)).
Looking at the overall dislocation trend, we observe a similarity between the LAGB and NTGB as the peak dislocation density coincides with void nucleation, evident from the drop in dislocation activity as shown in Figure \ref{fig:dislocation_temporal_profile}(b). 
However, for the IAGB, the decline is more gradual and plateaus after reaching the peak, following this a sharp decline is observed similar to the other two types of GB.

\begin{figure*}
\centering
	\begin{subfigure}[b]{0.8\textwidth}
		\centering
		\includegraphics[width=\linewidth]{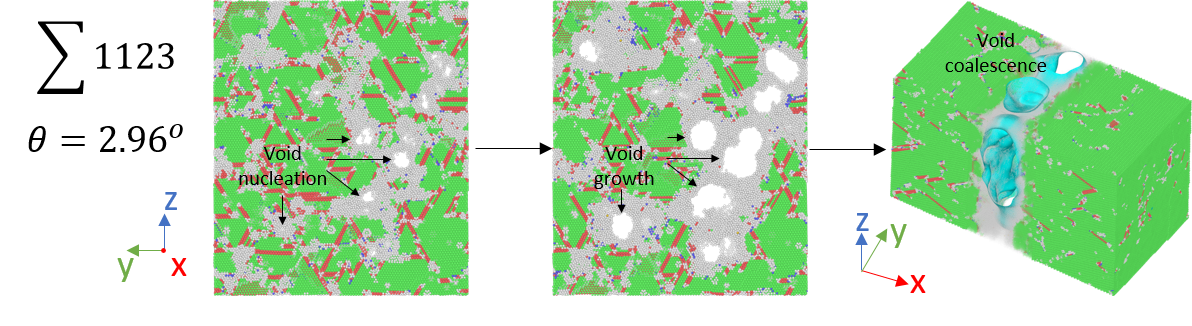}
		\caption{}
	\end{subfigure}
	\begin{subfigure}[b]{0.8\textwidth}
		\centering
		\includegraphics[width=\linewidth]{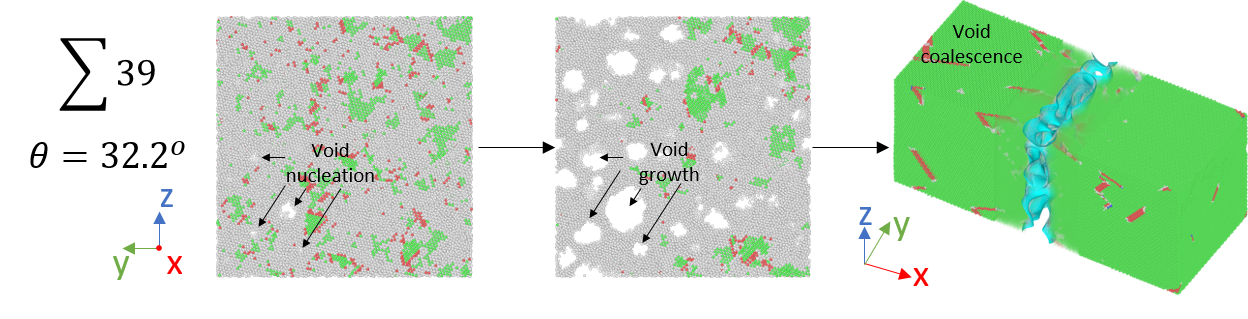}
		\caption{}
	\end{subfigure}
	\begin{subfigure}[b]{0.8\textwidth}
		\centering
		\includegraphics[width=\linewidth]{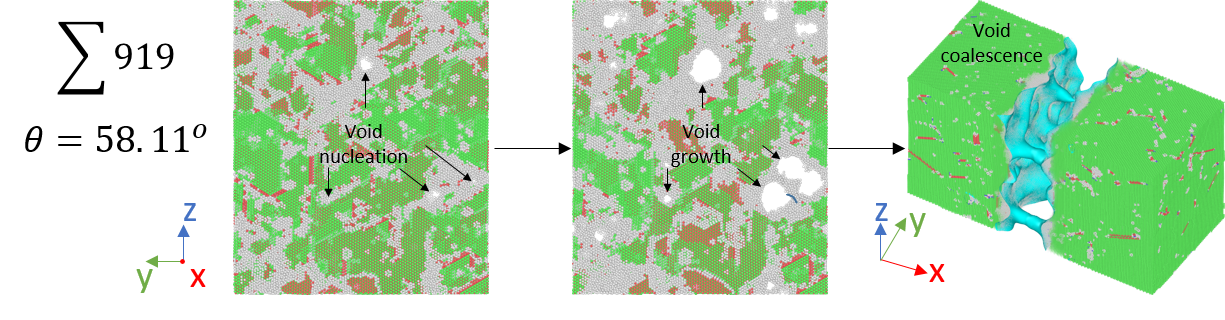}
		\caption{}
	\end{subfigure}
    \caption{Schematic of the spall failure for (a) LAGB ($\Sigma 1123$, $\theta = 2.96^\circ$), (b) IAGB ($\Sigma 39$, $\theta = 32.2^\circ$), and (c) NTGB ($\Sigma 919$, $\theta = 58.11^\circ$) at their respective $F_{\mathrm{abs}}^{\mathrm{spall}}$  where white, red, blue, and green represents defect, HCP, and BCC atoms, respectively. A surface mesh, based on the alpha method \cite{Stukowski2012}, is superimposed on the sample to highlight the coalescence of voids.} \label{fig:spall_failure}
\end{figure*}

The arrival of the tensile waves triggers spall failure, marked by void nucleation, growth, and coalescence, as discussed earlier and coinciding with the attainment of peak dislocation density.
Figure \ref{fig:spall_failure} illustrates the spall failure process, highlighting void nucleation and growth at the GB, with void coalescence occurring within a region of 50 nm in thickness.
Analyzing the LAGB and NTGB case, we can observe that voids tend to nucleate at weak points within the GB where due to the interaction of Shockley partials, atoms are displaced and tend to shear against each other \cite{Dongare2011,Pang2014}.
Once sufficient tensile stress is attained, atoms continue to displace, leading to the formation of voids and regions of displaced atoms.
As the voids expand, they interact with each other and coalescence.
Consequently, the presence of a LAGB or NTGB enables void nucleation due to the initial dislocation network and the promoted dislocation activity during the compression stage.
In the case of the IAGB, despite the GB acting as a barrier for dislocation activity between the two grains, void nucleation occurs at random locations due to initially displaced atoms inherent to the IAGB (as seen in Figure \ref{fig:GB_panel}(c)).
Subsequently, void growth and coalescence occurs during the tension stage.
As highlighted, the $F_{\mathrm{abs}}^{\mathrm{spall}}$ of IAGB is typically lower than  that of LAGB and NTGB (see Table \ref{tab:CSL_spall_strength}), and this could be attributed to the increase in defective atoms at the GB, which increases the likelihood of spallation due to the presence of more weak spots.

In summary, \textbf{photo-induced} shock loading, using the \textbf{two-pulse laser} method, was used to investigate the spall behaviour of (111) twist grain boundaries in Ni bicrystals by employing the $\ell$2T-MD model.
The different phenomena resulting from light-matter interaction revealed that the amount of deposited energy controls the type of laser-induced damage.
During the shock compression stage, the results highlighted the role of triple junction nodes, seen in certain low-angle GBs (LAGBs, $\theta\rightarrow 0^\circ$) and near-twin GBs (NTGBs, $\theta\rightarrow 60^\circ$).
The presence of these intersections promoted extensive dislocation growth and interaction.
However, for specific intermediate-angle GBs (IAGBs, $\theta\rightarrow 30^\circ$), the GB acts as a barrier, preventing the dislocation interaction between the two grains.
At the tension stage, it was observed that the misorientation angle has a significant influence on spall strength.
Generally, IAGBs exhibited better spall resistance, demonstrating a 25$\%$ improvement in spall strength across the considered misorientation angles, highlighting that GBs in Ni offer greater spall failure resistance compared to the bulk (e.g., $\sigma^{\mathrm{spall}}$ of $\Sigma1$ versus the Ni bi-crystals in Table \ref{tab:CSL_spall_strength}).
Failure analysis indicated that LAGB and NTGB tended to nucleate voids due to GB plasticity that took place at the compression stage, whereas IAGBs exhibited failure tendencies rooted in the inherent presence of defective atoms at the GB. 
This \textbf{photo-mechanical} approach can be further used to examine multi-layer metallic systems and other types of GB to provide more insights into material spall behaviour.
\section*{Acknowledgements}
We acknowledge the support of the New Frontiers in Research Fund 
(NFRFE-2019-01095) and the Natural Sciences and Engineering 
Research Council of Canada (NSERC) through the Discovery Grant 
and ALLRP~560447-2020 grants.
This research was supported through high-performance computational resources and services provided by Advanced Research Computing at the University of British Columbia and the Digital Research Alliance of Canada. 
We acknowledge the support of Canada Foundation for Innovation (CFI).
\section*{Conflict of Interest}
The authors have no conflicts to disclose.
\section*{Author Contributions}
\textbf{Mewael Isiet}: Methodology, Software, Validation, Formal analysis, Investigation, Data Curation, Visualization, Writing - Original Draft. \textbf{Mauricio Ponga}: Conceptualization, Methodology, Software, Resources, Project administration, Funding acquisition, Supervision, Writing - Review \& Editing.
\section*{Data availability}
The data that support the findings of this study are available from the corresponding author upon reasonable request.
\section*{Code availability}
The code developed to analytically calculate the thermoelastic stress can be found at \url{https://github.com/mewael-isiet/analytical-thermoelastic-wave}.
\section*{References}
\bibliography{aipsamp}

\begin{thebibliography}{46}%
\makeatletter
\providecommand \@ifxundefined [1]{%
 \@ifx{#1\undefined}
}%
\providecommand \@ifnum [1]{%
 \ifnum #1\expandafter \@firstoftwo
 \else \expandafter \@secondoftwo
 \fi
}%
\providecommand \@ifx [1]{%
 \ifx #1\expandafter \@firstoftwo
 \else \expandafter \@secondoftwo
 \fi
}%
\providecommand \natexlab [1]{#1}%
\providecommand \enquote  [1]{``#1''}%
\providecommand \bibnamefont  [1]{#1}%
\providecommand \bibfnamefont [1]{#1}%
\providecommand \citenamefont [1]{#1}%
\providecommand \href@noop [0]{\@secondoftwo}%
\providecommand \href [0]{\begingroup \@sanitize@url \@href}%
\providecommand \@href[1]{\@@startlink{#1}\@@href}%
\providecommand \@@href[1]{\endgroup#1\@@endlink}%
\providecommand \@sanitize@url [0]{\catcode `\\12\catcode `\$12\catcode
  `\&12\catcode `\#12\catcode `\^12\catcode `\_12\catcode `\%12\relax}%
\providecommand \@@startlink[1]{}%
\providecommand \@@endlink[0]{}%
\providecommand \url  [0]{\begingroup\@sanitize@url \@url }%
\providecommand \@url [1]{\endgroup\@href {#1}{\urlprefix }}%
\providecommand \urlprefix  [0]{URL }%
\providecommand \Eprint [0]{\href }%
\providecommand \doibase [0]{http://dx.doi.org/}%
\providecommand \selectlanguage [0]{\@gobble}%
\providecommand \bibinfo  [0]{\@secondoftwo}%
\providecommand \bibfield  [0]{\@secondoftwo}%
\providecommand \translation [1]{[#1]}%
\providecommand \BibitemOpen [0]{}%
\providecommand \bibitemStop [0]{}%
\providecommand \bibitemNoStop [0]{.\EOS\space}%
\providecommand \EOS [0]{\spacefactor3000\relax}%
\providecommand \BibitemShut  [1]{\csname bibitem#1\endcsname}%
\let\auto@bib@innerbib\@empty
\bibitem [{\citenamefont {de~Ress{\'e}guier}\ \emph {et~al.}(2017)\citenamefont
  {de~Ress{\'e}guier}, \citenamefont {Hemery}, \citenamefont {Lescoute},
  \citenamefont {Villechaise}, \citenamefont {Kanel},\ and\ \citenamefont
  {Razorenov}}]{deRessguier2017}%
  \BibitemOpen
  \bibfield  {author} {\bibinfo {author} {\bibfnamefont {T.}~\bibnamefont
  {de~Ress{\'e}guier}}, \bibinfo {author} {\bibfnamefont {S.}~\bibnamefont
  {Hemery}}, \bibinfo {author} {\bibfnamefont {E.}~\bibnamefont {Lescoute}},
  \bibinfo {author} {\bibfnamefont {P.}~\bibnamefont {Villechaise}}, \bibinfo
  {author} {\bibfnamefont {G.~I.}\ \bibnamefont {Kanel}}, \ and\ \bibinfo
  {author} {\bibfnamefont {S.~V.}\ \bibnamefont {Razorenov}},\ }\bibfield
  {title} {\enquote {\bibinfo {title} {Spall fracture and twinning in laser
  shock-loaded single-crystal magnesium},}\ }\href {\doibase 10.1063/1.4982352}
  {\bibfield  {journal} {\bibinfo  {journal} {Journal of Applied Physics}\
  }\textbf {\bibinfo {volume} {121}},\ \bibinfo {pages} {165104} (\bibinfo
  {year} {2017})}\BibitemShut {NoStop}%
\bibitem [{\citenamefont {Kanel}(2010)}]{Kanel2010}%
  \BibitemOpen
  \bibfield  {author} {\bibinfo {author} {\bibfnamefont {G.~I.}\ \bibnamefont
  {Kanel}},\ }\bibfield  {title} {\enquote {\bibinfo {title} {Spall fracture:
  methodological aspects, mechanisms and governing factors},}\ }\href {\doibase
  10.1007/s10704-009-9438-0} {\bibfield  {journal} {\bibinfo  {journal}
  {International Journal of Fracture}\ }\textbf {\bibinfo {volume} {163}},\
  \bibinfo {pages} {173--191} (\bibinfo {year} {2010})}\BibitemShut {NoStop}%
\bibitem [{\citenamefont {Echeverria}\ \emph {et~al.}(2021)\citenamefont
  {Echeverria}, \citenamefont {Galitskiy}, \citenamefont {Mishra},
  \citenamefont {Dingreville},\ and\ \citenamefont {Dongare}}]{Echeverria2021}%
  \BibitemOpen
  \bibfield  {author} {\bibinfo {author} {\bibfnamefont {M.~J.}\ \bibnamefont
  {Echeverria}}, \bibinfo {author} {\bibfnamefont {S.}~\bibnamefont
  {Galitskiy}}, \bibinfo {author} {\bibfnamefont {A.}~\bibnamefont {Mishra}},
  \bibinfo {author} {\bibfnamefont {R.}~\bibnamefont {Dingreville}}, \ and\
  \bibinfo {author} {\bibfnamefont {A.~M.}\ \bibnamefont {Dongare}},\
  }\bibfield  {title} {\enquote {\bibinfo {title} {Understanding the plasticity
  contributions during laser-shock loading and spall failure of cu
  microstructures at the atomic scales},}\ }\href {\doibase
  10.1016/j.commatsci.2021.110668} {\bibfield  {journal} {\bibinfo  {journal}
  {Computational Materials Science}\ }\textbf {\bibinfo {volume} {198}},\
  \bibinfo {pages} {110668} (\bibinfo {year} {2021})}\BibitemShut {NoStop}%
\bibitem [{\citenamefont {Isiet}\ \emph {et~al.}(2024)\citenamefont {Isiet},
  \citenamefont {Xiao}, \citenamefont {Dadap}, \citenamefont {Ye},\ and\
  \citenamefont {Ponga}}]{1st_paper}%
  \BibitemOpen
  \bibfield  {author} {\bibinfo {author} {\bibfnamefont {M.}~\bibnamefont
  {Isiet}}, \bibinfo {author} {\bibfnamefont {Y.}~\bibnamefont {Xiao}},
  \bibinfo {author} {\bibfnamefont {J.~I.}\ \bibnamefont {Dadap}}, \bibinfo
  {author} {\bibfnamefont {Z.}~\bibnamefont {Ye}}, \ and\ \bibinfo {author}
  {\bibfnamefont {M.}~\bibnamefont {Ponga}},\ }\bibfield  {title} {\enquote
  {\bibinfo {title} {Femtosecond two-pulse laser approach for spall failure in
  thin foils},}\ }\href@noop {} {\bibfield  {journal} {\bibinfo  {journal}
  {arXiv preprint arXiv:2412.04762}\ } (\bibinfo {year} {2024})}\BibitemShut
  {NoStop}%
\bibitem [{\citenamefont {Remington}\ \emph {et~al.}(2018)\citenamefont
  {Remington}, \citenamefont {Hahn}, \citenamefont {Zhao}, \citenamefont
  {Flanagan}, \citenamefont {Mertens}, \citenamefont {Sabbaghianrad},
  \citenamefont {Langdon}, \citenamefont {Wehrenberg}, \citenamefont {Maddox},
  \citenamefont {Swift}, \citenamefont {Remington}, \citenamefont {Chawla},\
  and\ \citenamefont {Meyers}}]{Remington2018}%
  \BibitemOpen
  \bibfield  {author} {\bibinfo {author} {\bibfnamefont {T.}~\bibnamefont
  {Remington}}, \bibinfo {author} {\bibfnamefont {E.}~\bibnamefont {Hahn}},
  \bibinfo {author} {\bibfnamefont {S.}~\bibnamefont {Zhao}}, \bibinfo {author}
  {\bibfnamefont {R.}~\bibnamefont {Flanagan}}, \bibinfo {author}
  {\bibfnamefont {J.}~\bibnamefont {Mertens}}, \bibinfo {author} {\bibfnamefont
  {S.}~\bibnamefont {Sabbaghianrad}}, \bibinfo {author} {\bibfnamefont
  {T.}~\bibnamefont {Langdon}}, \bibinfo {author} {\bibfnamefont
  {C.}~\bibnamefont {Wehrenberg}}, \bibinfo {author} {\bibfnamefont
  {B.}~\bibnamefont {Maddox}}, \bibinfo {author} {\bibfnamefont
  {D.}~\bibnamefont {Swift}}, \bibinfo {author} {\bibfnamefont
  {B.}~\bibnamefont {Remington}}, \bibinfo {author} {\bibfnamefont
  {N.}~\bibnamefont {Chawla}}, \ and\ \bibinfo {author} {\bibfnamefont
  {M.}~\bibnamefont {Meyers}},\ }\bibfield  {title} {\enquote {\bibinfo {title}
  {Spall strength dependence on grain size and strain rate in tantalum},}\
  }\href {\doibase 10.1016/j.actamat.2018.07.048} {\bibfield  {journal}
  {\bibinfo  {journal} {Acta Materialia}\ }\textbf {\bibinfo {volume} {158}},\
  \bibinfo {pages} {313--329} (\bibinfo {year} {2018})}\BibitemShut {NoStop}%
\bibitem [{\citenamefont {Ariza}\ \emph {et~al.}(2011)\citenamefont {Ariza},
  \citenamefont {Romero}, \citenamefont {Ponga},\ and\ \citenamefont
  {Ortiz}}]{Ariza2011}%
  \BibitemOpen
  \bibfield  {author} {\bibinfo {author} {\bibfnamefont {M.~P.}\ \bibnamefont
  {Ariza}}, \bibinfo {author} {\bibfnamefont {I.}~\bibnamefont {Romero}},
  \bibinfo {author} {\bibfnamefont {M.}~\bibnamefont {Ponga}}, \ and\ \bibinfo
  {author} {\bibfnamefont {M.}~\bibnamefont {Ortiz}},\ }\bibfield  {title}
  {\enquote {\bibinfo {title} {Hotqc simulation of nanovoid growth under
  tension in copper},}\ }\href {\doibase 10.1007/s10704-011-9660-4} {\bibfield
  {journal} {\bibinfo  {journal} {International Journal of Fracture}\ }\textbf
  {\bibinfo {volume} {174}},\ \bibinfo {pages} {75–85} (\bibinfo {year}
  {2011})}\BibitemShut {NoStop}%
\bibitem [{\citenamefont {Ponga}\ \emph {et~al.}(2016)\citenamefont {Ponga},
  \citenamefont {Ramabathiran}, \citenamefont {Bhattacharya},\ and\
  \citenamefont {Ortiz}}]{Ponga2016}%
  \BibitemOpen
  \bibfield  {author} {\bibinfo {author} {\bibfnamefont {M.}~\bibnamefont
  {Ponga}}, \bibinfo {author} {\bibfnamefont {A.~A.}\ \bibnamefont
  {Ramabathiran}}, \bibinfo {author} {\bibfnamefont {K.}~\bibnamefont
  {Bhattacharya}}, \ and\ \bibinfo {author} {\bibfnamefont {M.}~\bibnamefont
  {Ortiz}},\ }\bibfield  {title} {\enquote {\bibinfo {title} {Dynamic behavior
  of nano-voids in magnesium under hydrostatic tensile stress},}\ }\href
  {\doibase 10.1088/0965-0393/24/6/065003} {\bibfield  {journal} {\bibinfo
  {journal} {Modelling and Simulation in Materials Science and Engineering}\
  }\textbf {\bibinfo {volume} {24}},\ \bibinfo {pages} {065003} (\bibinfo
  {year} {2016})}\BibitemShut {NoStop}%
\bibitem [{\citenamefont {Grégoire}\ and\ \citenamefont
  {Ponga}(2017)}]{Grgoire2017}%
  \BibitemOpen
  \bibfield  {author} {\bibinfo {author} {\bibfnamefont {C.}~\bibnamefont
  {Grégoire}}\ and\ \bibinfo {author} {\bibfnamefont {M.}~\bibnamefont
  {Ponga}},\ }\bibfield  {title} {\enquote {\bibinfo {title} {Nanovoid failure
  in magnesium under dynamic loads},}\ }\href {\doibase
  10.1016/j.actamat.2017.05.016} {\bibfield  {journal} {\bibinfo  {journal}
  {Acta Materialia}\ }\textbf {\bibinfo {volume} {134}},\ \bibinfo {pages}
  {360–374} (\bibinfo {year} {2017})}\BibitemShut {NoStop}%
\bibitem [{\citenamefont {Bringa}, \citenamefont {Traiviratana},\ and\
  \citenamefont {Meyers}(2010)}]{Bringa2010}%
  \BibitemOpen
  \bibfield  {author} {\bibinfo {author} {\bibfnamefont {E.~M.}\ \bibnamefont
  {Bringa}}, \bibinfo {author} {\bibfnamefont {S.}~\bibnamefont
  {Traiviratana}}, \ and\ \bibinfo {author} {\bibfnamefont {M.~A.}\
  \bibnamefont {Meyers}},\ }\bibfield  {title} {\enquote {\bibinfo {title}
  {Void initiation in fcc metals: Effect of loading orientation and
  nanocrystalline effects},}\ }\href {\doibase 10.1016/j.actamat.2010.04.043}
  {\bibfield  {journal} {\bibinfo  {journal} {Acta Materialia}\ }\textbf
  {\bibinfo {volume} {58}},\ \bibinfo {pages} {4458–4477} (\bibinfo {year}
  {2010})}\BibitemShut {NoStop}%
\bibitem [{\citenamefont {Remington}\ \emph {et~al.}(2006)\citenamefont
  {Remington}, \citenamefont {Allen}, \citenamefont {Bringa}, \citenamefont
  {Hawreliak}, \citenamefont {Ho}, \citenamefont {Lorenz}, \citenamefont
  {Lorenzana}, \citenamefont {McNaney}, \citenamefont {Meyers}, \citenamefont
  {Pollaine}, \citenamefont {Rosolankova}, \citenamefont {Sadik}, \citenamefont
  {Schneider}, \citenamefont {Swift}, \citenamefont {Wark},\ and\ \citenamefont
  {Yaakobi}}]{Remington2006}%
  \BibitemOpen
  \bibfield  {author} {\bibinfo {author} {\bibfnamefont {B.~A.}\ \bibnamefont
  {Remington}}, \bibinfo {author} {\bibfnamefont {P.}~\bibnamefont {Allen}},
  \bibinfo {author} {\bibfnamefont {E.~M.}\ \bibnamefont {Bringa}}, \bibinfo
  {author} {\bibfnamefont {J.}~\bibnamefont {Hawreliak}}, \bibinfo {author}
  {\bibfnamefont {D.}~\bibnamefont {Ho}}, \bibinfo {author} {\bibfnamefont
  {K.~T.}\ \bibnamefont {Lorenz}}, \bibinfo {author} {\bibfnamefont
  {H.}~\bibnamefont {Lorenzana}}, \bibinfo {author} {\bibfnamefont {J.~M.}\
  \bibnamefont {McNaney}}, \bibinfo {author} {\bibfnamefont {M.~A.}\
  \bibnamefont {Meyers}}, \bibinfo {author} {\bibfnamefont {S.~W.}\
  \bibnamefont {Pollaine}}, \bibinfo {author} {\bibfnamefont {K.}~\bibnamefont
  {Rosolankova}}, \bibinfo {author} {\bibfnamefont {B.}~\bibnamefont {Sadik}},
  \bibinfo {author} {\bibfnamefont {M.~S.}\ \bibnamefont {Schneider}}, \bibinfo
  {author} {\bibfnamefont {D.}~\bibnamefont {Swift}}, \bibinfo {author}
  {\bibfnamefont {J.}~\bibnamefont {Wark}}, \ and\ \bibinfo {author}
  {\bibfnamefont {B.}~\bibnamefont {Yaakobi}},\ }\bibfield  {title} {\enquote
  {\bibinfo {title} {Material dynamics under extreme conditions of pressure and
  strain rate},}\ }\href {\doibase 10.1179/174328406x91069} {\bibfield
  {journal} {\bibinfo  {journal} {Materials Science and Technology}\ }\textbf
  {\bibinfo {volume} {22}},\ \bibinfo {pages} {474–488} (\bibinfo {year}
  {2006})}\BibitemShut {NoStop}%
\bibitem [{\citenamefont {Kumar}, \citenamefont {Swygenhoven},\ and\
  \citenamefont {Suresh}(2003)}]{Kumar2003_1}%
  \BibitemOpen
  \bibfield  {author} {\bibinfo {author} {\bibfnamefont {K.}~\bibnamefont
  {Kumar}}, \bibinfo {author} {\bibfnamefont {H.~V.}\ \bibnamefont
  {Swygenhoven}}, \ and\ \bibinfo {author} {\bibfnamefont {S.}~\bibnamefont
  {Suresh}},\ }\bibfield  {title} {\enquote {\bibinfo {title} {Mechanical
  behavior of nanocrystalline metals and alloys},}\ }\href {\doibase
  10.1016/j.actamat.2003.08.032} {\bibfield  {journal} {\bibinfo  {journal}
  {Acta Materialia}\ }\textbf {\bibinfo {volume} {51}},\ \bibinfo {pages}
  {5743--5774} (\bibinfo {year} {2003})}\BibitemShut {NoStop}%
\bibitem [{\citenamefont {Fensin}\ \emph {et~al.}(2014)\citenamefont {Fensin},
  \citenamefont {Cerreta}, \citenamefont {III},\ and\ \citenamefont
  {Valone}}]{Fensin2014}%
  \BibitemOpen
  \bibfield  {author} {\bibinfo {author} {\bibfnamefont {S.~J.}\ \bibnamefont
  {Fensin}}, \bibinfo {author} {\bibfnamefont {E.~K.}\ \bibnamefont {Cerreta}},
  \bibinfo {author} {\bibfnamefont {G.~T.~G.}\ \bibnamefont {III}}, \ and\
  \bibinfo {author} {\bibfnamefont {S.~M.}\ \bibnamefont {Valone}},\ }\bibfield
   {title} {\enquote {\bibinfo {title} {Why are some interfaces in materials
  stronger than others?}}\ }\href {\doibase 10.1038/srep05461} {\bibfield
  {journal} {\bibinfo  {journal} {Scientific Reports}\ }\textbf {\bibinfo
  {volume} {4}} (\bibinfo {year} {2014}),\ 10.1038/srep05461}\BibitemShut
  {NoStop}%
\bibitem [{\citenamefont {Cerreta}\ \emph {et~al.}(2012)\citenamefont
  {Cerreta}, \citenamefont {Escobedo}, \citenamefont {Perez-Bergquist},
  \citenamefont {Koller}, \citenamefont {Trujillo}, \citenamefont {III},
  \citenamefont {Brandl},\ and\ \citenamefont {Germann}}]{Cerreta2012}%
  \BibitemOpen
  \bibfield  {author} {\bibinfo {author} {\bibfnamefont {E.}~\bibnamefont
  {Cerreta}}, \bibinfo {author} {\bibfnamefont {J.}~\bibnamefont {Escobedo}},
  \bibinfo {author} {\bibfnamefont {A.}~\bibnamefont {Perez-Bergquist}},
  \bibinfo {author} {\bibfnamefont {D.}~\bibnamefont {Koller}}, \bibinfo
  {author} {\bibfnamefont {C.}~\bibnamefont {Trujillo}}, \bibinfo {author}
  {\bibfnamefont {G.~G.}\ \bibnamefont {III}}, \bibinfo {author} {\bibfnamefont
  {C.}~\bibnamefont {Brandl}}, \ and\ \bibinfo {author} {\bibfnamefont
  {T.}~\bibnamefont {Germann}},\ }\bibfield  {title} {\enquote {\bibinfo
  {title} {Early stage dynamic damage and the role of grain boundary type},}\
  }\href {\doibase 10.1016/j.scriptamat.2012.01.051} {\bibfield  {journal}
  {\bibinfo  {journal} {Scripta Materialia}\ }\textbf {\bibinfo {volume}
  {66}},\ \bibinfo {pages} {638--641} (\bibinfo {year} {2012})}\BibitemShut
  {NoStop}%
\bibitem [{\citenamefont {Wayne}\ \emph {et~al.}(2010)\citenamefont {Wayne},
  \citenamefont {Krishnan}, \citenamefont {DiGiacomo}, \citenamefont {Kovvali},
  \citenamefont {Peralta}, \citenamefont {Luo}, \citenamefont {Greenfield},
  \citenamefont {Byler}, \citenamefont {Paisley}, \citenamefont {McClellan},
  \citenamefont {Koskelo},\ and\ \citenamefont {Dickerson}}]{Wayne2010}%
  \BibitemOpen
  \bibfield  {author} {\bibinfo {author} {\bibfnamefont {L.}~\bibnamefont
  {Wayne}}, \bibinfo {author} {\bibfnamefont {K.}~\bibnamefont {Krishnan}},
  \bibinfo {author} {\bibfnamefont {S.}~\bibnamefont {DiGiacomo}}, \bibinfo
  {author} {\bibfnamefont {N.}~\bibnamefont {Kovvali}}, \bibinfo {author}
  {\bibfnamefont {P.}~\bibnamefont {Peralta}}, \bibinfo {author} {\bibfnamefont
  {S.}~\bibnamefont {Luo}}, \bibinfo {author} {\bibfnamefont {S.}~\bibnamefont
  {Greenfield}}, \bibinfo {author} {\bibfnamefont {D.}~\bibnamefont {Byler}},
  \bibinfo {author} {\bibfnamefont {D.}~\bibnamefont {Paisley}}, \bibinfo
  {author} {\bibfnamefont {K.}~\bibnamefont {McClellan}}, \bibinfo {author}
  {\bibfnamefont {A.}~\bibnamefont {Koskelo}}, \ and\ \bibinfo {author}
  {\bibfnamefont {R.}~\bibnamefont {Dickerson}},\ }\bibfield  {title} {\enquote
  {\bibinfo {title} {Statistics of weak grain boundaries for spall damage in
  polycrystalline copper},}\ }\href {\doibase 10.1016/j.scriptamat.2010.08.003}
  {\bibfield  {journal} {\bibinfo  {journal} {Scripta Materialia}\ }\textbf
  {\bibinfo {volume} {63}},\ \bibinfo {pages} {1065--1068} (\bibinfo {year}
  {2010})}\BibitemShut {NoStop}%
\bibitem [{\citenamefont {Escobedo}\ \emph {et~al.}(2011)\citenamefont
  {Escobedo}, \citenamefont {Dennis-Koller}, \citenamefont {Cerreta},
  \citenamefont {Patterson}, \citenamefont {Bronkhorst}, \citenamefont
  {Hansen}, \citenamefont {Tonks},\ and\ \citenamefont
  {Lebensohn}}]{Escobedo2011}%
  \BibitemOpen
  \bibfield  {author} {\bibinfo {author} {\bibfnamefont {J.~P.}\ \bibnamefont
  {Escobedo}}, \bibinfo {author} {\bibfnamefont {D.}~\bibnamefont
  {Dennis-Koller}}, \bibinfo {author} {\bibfnamefont {E.~K.}\ \bibnamefont
  {Cerreta}}, \bibinfo {author} {\bibfnamefont {B.~M.}\ \bibnamefont
  {Patterson}}, \bibinfo {author} {\bibfnamefont {C.~A.}\ \bibnamefont
  {Bronkhorst}}, \bibinfo {author} {\bibfnamefont {B.~L.}\ \bibnamefont
  {Hansen}}, \bibinfo {author} {\bibfnamefont {D.}~\bibnamefont {Tonks}}, \
  and\ \bibinfo {author} {\bibfnamefont {R.~A.}\ \bibnamefont {Lebensohn}},\
  }\bibfield  {title} {\enquote {\bibinfo {title} {Effects of grain size and
  boundary structure on the dynamic tensile response of copper},}\ }\href
  {\doibase 10.1063/1.3607294} {\bibfield  {journal} {\bibinfo  {journal}
  {Journal of Applied Physics}\ }\textbf {\bibinfo {volume} {110}},\ \bibinfo
  {pages} {033513} (\bibinfo {year} {2011})}\BibitemShut {NoStop}%
\bibitem [{\citenamefont {Escobedo}, \citenamefont {Cerreta},\ and\
  \citenamefont {Dennis-Koller}(2013)}]{Escobedo2013}%
  \BibitemOpen
  \bibfield  {author} {\bibinfo {author} {\bibfnamefont {J.~P.}\ \bibnamefont
  {Escobedo}}, \bibinfo {author} {\bibfnamefont {E.~K.}\ \bibnamefont
  {Cerreta}}, \ and\ \bibinfo {author} {\bibfnamefont {D.}~\bibnamefont
  {Dennis-Koller}},\ }\bibfield  {title} {\enquote {\bibinfo {title} {Effect of
  crystalline structure on intergranular failure during shock loading},}\
  }\href {\doibase 10.1007/s11837-013-0798-6} {\bibfield  {journal} {\bibinfo
  {journal} {{JOM}}\ }\textbf {\bibinfo {volume} {66}},\ \bibinfo {pages}
  {156--164} (\bibinfo {year} {2013})}\BibitemShut {NoStop}%
\bibitem [{\citenamefont {Euser}\ \emph {et~al.}(2023)\citenamefont {Euser},
  \citenamefont {Jones}, \citenamefont {Martinez}, \citenamefont {Valdez},
  \citenamefont {Trujillo}, \citenamefont {Cady},\ and\ \citenamefont
  {Fensin}}]{Euser2023}%
  \BibitemOpen
  \bibfield  {author} {\bibinfo {author} {\bibfnamefont {V.}~\bibnamefont
  {Euser}}, \bibinfo {author} {\bibfnamefont {D.}~\bibnamefont {Jones}},
  \bibinfo {author} {\bibfnamefont {D.}~\bibnamefont {Martinez}}, \bibinfo
  {author} {\bibfnamefont {J.}~\bibnamefont {Valdez}}, \bibinfo {author}
  {\bibfnamefont {C.}~\bibnamefont {Trujillo}}, \bibinfo {author}
  {\bibfnamefont {C.}~\bibnamefont {Cady}}, \ and\ \bibinfo {author}
  {\bibfnamefont {S.}~\bibnamefont {Fensin}},\ }\bibfield  {title} {\enquote
  {\bibinfo {title} {The effect of microstructure on the dynamic shock response
  of 1045 steel},}\ }\href {\doibase 10.1016/j.actamat.2023.118874} {\bibfield
  {journal} {\bibinfo  {journal} {Acta Materialia}\ }\textbf {\bibinfo {volume}
  {250}},\ \bibinfo {pages} {118874} (\bibinfo {year} {2023})}\BibitemShut
  {NoStop}%
\bibitem [{\citenamefont {Konnur}, \citenamefont {Reddy},\ and\ \citenamefont
  {Pal}(2021)}]{Konnur2021}%
  \BibitemOpen
  \bibfield  {author} {\bibinfo {author} {\bibfnamefont {T.}~\bibnamefont
  {Konnur}}, \bibinfo {author} {\bibfnamefont {K.~V.}\ \bibnamefont {Reddy}}, \
  and\ \bibinfo {author} {\bibfnamefont {S.}~\bibnamefont {Pal}},\ }\bibfield
  {title} {\enquote {\bibinfo {title} {Effect of variation in inclination angle
  of $\sum5$ tilt grain boundary on the shock response of ni bicrystals},}\
  }\href {\doibase 10.1007/s00339-021-04502-z} {\bibfield  {journal} {\bibinfo
  {journal} {Applied Physics A}\ }\textbf {\bibinfo {volume} {127}} (\bibinfo
  {year} {2021}),\ 10.1007/s00339-021-04502-z}\BibitemShut {NoStop}%
\bibitem [{\citenamefont {Zhu}\ \emph {et~al.}(2022)\citenamefont {Zhu},
  \citenamefont {Hu}, \citenamefont {Huang}, \citenamefont {Wang},
  \citenamefont {Luo},\ and\ \citenamefont {Shen}}]{Zhu2022}%
  \BibitemOpen
  \bibfield  {author} {\bibinfo {author} {\bibfnamefont {Y.}~\bibnamefont
  {Zhu}}, \bibinfo {author} {\bibfnamefont {J.}~\bibnamefont {Hu}}, \bibinfo
  {author} {\bibfnamefont {S.}~\bibnamefont {Huang}}, \bibinfo {author}
  {\bibfnamefont {J.}~\bibnamefont {Wang}}, \bibinfo {author} {\bibfnamefont
  {G.}~\bibnamefont {Luo}}, \ and\ \bibinfo {author} {\bibfnamefont
  {Q.}~\bibnamefont {Shen}},\ }\bibfield  {title} {\enquote {\bibinfo {title}
  {Molecular dynamics simulation on spallation of [111] cu/ni nano-multilayers:
  Voids evolution under different shock pulse duration},}\ }\href {\doibase
  10.1016/j.commatsci.2021.110923} {\bibfield  {journal} {\bibinfo  {journal}
  {Computational Materials Science}\ }\textbf {\bibinfo {volume} {202}},\
  \bibinfo {pages} {110923} (\bibinfo {year} {2022})}\BibitemShut {NoStop}%
\bibitem [{\citenamefont {Long}\ \emph {et~al.}(2020)\citenamefont {Long},
  \citenamefont {Liu}, \citenamefont {Zhang}, \citenamefont {Peng},\ and\
  \citenamefont {Wang}}]{Long2020}%
  \BibitemOpen
  \bibfield  {author} {\bibinfo {author} {\bibfnamefont {X.}~\bibnamefont
  {Long}}, \bibinfo {author} {\bibfnamefont {X.}~\bibnamefont {Liu}}, \bibinfo
  {author} {\bibfnamefont {W.}~\bibnamefont {Zhang}}, \bibinfo {author}
  {\bibfnamefont {Y.}~\bibnamefont {Peng}}, \ and\ \bibinfo {author}
  {\bibfnamefont {G.}~\bibnamefont {Wang}},\ }\bibfield  {title} {\enquote
  {\bibinfo {title} {Shock deformation and spallation of cu bicrystals with (1
  1 1) twist grain boundaries},}\ }\href {\doibase
  10.1016/j.commatsci.2019.109411} {\bibfield  {journal} {\bibinfo  {journal}
  {Computational Materials Science}\ }\textbf {\bibinfo {volume} {173}},\
  \bibinfo {pages} {109411} (\bibinfo {year} {2020})}\BibitemShut {NoStop}%
\bibitem [{\citenamefont {Ponga}\ and\ \citenamefont {Sun}(2018)}]{Ponga2018}%
  \BibitemOpen
  \bibfield  {author} {\bibinfo {author} {\bibfnamefont {M.}~\bibnamefont
  {Ponga}}\ and\ \bibinfo {author} {\bibfnamefont {D.}~\bibnamefont {Sun}},\
  }\bibfield  {title} {\enquote {\bibinfo {title} {A unified framework for heat
  and mass transport at the atomic scale},}\ }\href {\doibase
  10.1088/1361-651x/aaaf94} {\bibfield  {journal} {\bibinfo  {journal}
  {Modelling and Simulation in Materials Science and Engineering}\ }\textbf
  {\bibinfo {volume} {26}},\ \bibinfo {pages} {035014} (\bibinfo {year}
  {2018})}\BibitemShut {NoStop}%
\bibitem [{\citenamefont {Ullah}\ and\ \citenamefont
  {Ponga}(2019)}]{Ullah2019}%
  \BibitemOpen
  \bibfield  {author} {\bibinfo {author} {\bibfnamefont {M.~W.}\ \bibnamefont
  {Ullah}}\ and\ \bibinfo {author} {\bibfnamefont {M.}~\bibnamefont {Ponga}},\
  }\bibfield  {title} {\enquote {\bibinfo {title} {A new approach for
  electronic heat conduction in molecular dynamics simulations},}\ }\href
  {\doibase 10.1088/1361-651x/ab309f} {\bibfield  {journal} {\bibinfo
  {journal} {Modelling and Simulation in Materials Science and Engineering}\
  }\textbf {\bibinfo {volume} {27}},\ \bibinfo {pages} {075008} (\bibinfo
  {year} {2019})}\BibitemShut {NoStop}%
\bibitem [{\citenamefont {Hendy}\ and\ \citenamefont
  {Ponga}(2023)}]{Hendy2023}%
  \BibitemOpen
  \bibfield  {author} {\bibinfo {author} {\bibfnamefont {M.}~\bibnamefont
  {Hendy}}\ and\ \bibinfo {author} {\bibfnamefont {M.}~\bibnamefont {Ponga}},\
  }\bibfield  {title} {\enquote {\bibinfo {title} {A multiscale and
  multiphysics framework to simulate radiation damage in nano-crystalline
  materials},}\ }\href {\doibase 10.1016/j.jnucmat.2023.154347} {\bibfield
  {journal} {\bibinfo  {journal} {Journal of Nuclear Materials}\ }\textbf
  {\bibinfo {volume} {578}},\ \bibinfo {pages} {154347} (\bibinfo {year}
  {2023})}\BibitemShut {NoStop}%
\bibitem [{\citenamefont {Plimpton}(1995)}]{Plimpton1995}%
  \BibitemOpen
  \bibfield  {author} {\bibinfo {author} {\bibfnamefont {S.}~\bibnamefont
  {Plimpton}},\ }\bibfield  {title} {\enquote {\bibinfo {title} {Fast parallel
  algorithms for short-range molecular dynamics},}\ }\href {\doibase
  10.1006/jcph.1995.1039} {\bibfield  {journal} {\bibinfo  {journal} {Journal
  of Computational Physics}\ }\textbf {\bibinfo {volume} {117}},\ \bibinfo
  {pages} {1--19} (\bibinfo {year} {1995})}\BibitemShut {NoStop}%
\bibitem [{\citenamefont {Thompson}\ \emph {et~al.}(2022)\citenamefont
  {Thompson}, \citenamefont {Aktulga}, \citenamefont {Berger}, \citenamefont
  {Bolintineanu}, \citenamefont {Brown}, \citenamefont {Crozier}, \citenamefont
  {in~'t Veld}, \citenamefont {Kohlmeyer}, \citenamefont {Moore}, \citenamefont
  {Nguyen}, \citenamefont {Shan}, \citenamefont {Stevens}, \citenamefont
  {Tranchida}, \citenamefont {Trott},\ and\ \citenamefont {Plimpton}}]{LAMMPS}%
  \BibitemOpen
  \bibfield  {author} {\bibinfo {author} {\bibfnamefont {A.~P.}\ \bibnamefont
  {Thompson}}, \bibinfo {author} {\bibfnamefont {H.~M.}\ \bibnamefont
  {Aktulga}}, \bibinfo {author} {\bibfnamefont {R.}~\bibnamefont {Berger}},
  \bibinfo {author} {\bibfnamefont {D.~S.}\ \bibnamefont {Bolintineanu}},
  \bibinfo {author} {\bibfnamefont {W.~M.}\ \bibnamefont {Brown}}, \bibinfo
  {author} {\bibfnamefont {P.~S.}\ \bibnamefont {Crozier}}, \bibinfo {author}
  {\bibfnamefont {P.~J.}\ \bibnamefont {in~'t Veld}}, \bibinfo {author}
  {\bibfnamefont {A.}~\bibnamefont {Kohlmeyer}}, \bibinfo {author}
  {\bibfnamefont {S.~G.}\ \bibnamefont {Moore}}, \bibinfo {author}
  {\bibfnamefont {T.~D.}\ \bibnamefont {Nguyen}}, \bibinfo {author}
  {\bibfnamefont {R.}~\bibnamefont {Shan}}, \bibinfo {author} {\bibfnamefont
  {M.~J.}\ \bibnamefont {Stevens}}, \bibinfo {author} {\bibfnamefont
  {J.}~\bibnamefont {Tranchida}}, \bibinfo {author} {\bibfnamefont
  {C.}~\bibnamefont {Trott}}, \ and\ \bibinfo {author} {\bibfnamefont {S.~J.}\
  \bibnamefont {Plimpton}},\ }\bibfield  {title} {\enquote {\bibinfo {title}
  {{LAMMPS} - a flexible simulation tool for particle-based materials modeling
  at the atomic, meso, and continuum scales},}\ }\href {\doibase
  10.1016/j.cpc.2021.108171} {\bibfield  {journal} {\bibinfo  {journal} {Comp.
  Phys. Comm.}\ }\textbf {\bibinfo {volume} {271}},\ \bibinfo {pages} {108171}
  (\bibinfo {year} {2022})}\BibitemShut {NoStop}%
\bibitem [{\citenamefont {Daw}\ and\ \citenamefont {Baskes}(1984)}]{Daw1984}%
  \BibitemOpen
  \bibfield  {author} {\bibinfo {author} {\bibfnamefont {M.~S.}\ \bibnamefont
  {Daw}}\ and\ \bibinfo {author} {\bibfnamefont {M.~I.}\ \bibnamefont
  {Baskes}},\ }\bibfield  {title} {\enquote {\bibinfo {title} {Embedded-atom
  method: Derivation and application to impurities, surfaces, and other defects
  in metals},}\ }\href {\doibase 10.1103/physrevb.29.6443} {\bibfield
  {journal} {\bibinfo  {journal} {Physical Review B}\ }\textbf {\bibinfo
  {volume} {29}},\ \bibinfo {pages} {6443--6453} (\bibinfo {year}
  {1984})}\BibitemShut {NoStop}%
\bibitem [{\citenamefont {Mishin}(2004)}]{Mishin2004}%
  \BibitemOpen
  \bibfield  {author} {\bibinfo {author} {\bibfnamefont {Y.}~\bibnamefont
  {Mishin}},\ }\bibfield  {title} {\enquote {\bibinfo {title} {Atomistic
  modeling of the $\gamma$ and $\gamma$'-phases of the ni--al system},}\ }\href
  {\doibase 10.1016/j.actamat.2003.11.026} {\bibfield  {journal} {\bibinfo
  {journal} {Acta Materialia}\ }\textbf {\bibinfo {volume} {52}},\ \bibinfo
  {pages} {1451--1467} (\bibinfo {year} {2004})}\BibitemShut {NoStop}%
\bibitem [{\citenamefont {Dai}, \citenamefont {Xiang},\ and\ \citenamefont
  {Srolovitz}(2014)}]{Dai2014}%
  \BibitemOpen
  \bibfield  {author} {\bibinfo {author} {\bibfnamefont {S.}~\bibnamefont
  {Dai}}, \bibinfo {author} {\bibfnamefont {Y.}~\bibnamefont {Xiang}}, \ and\
  \bibinfo {author} {\bibfnamefont {D.~J.}\ \bibnamefont {Srolovitz}},\
  }\bibfield  {title} {\enquote {\bibinfo {title} {Atomistic, generalized
  peierls{\textendash}nabarro and analytical models for (111) twist boundaries
  in al, cu and ni for all twist angles},}\ }\href {\doibase
  10.1016/j.actamat.2014.01.022} {\bibfield  {journal} {\bibinfo  {journal}
  {Acta Materialia}\ }\textbf {\bibinfo {volume} {69}},\ \bibinfo {pages}
  {162--174} (\bibinfo {year} {2014})}\BibitemShut {NoStop}%
\bibitem [{\citenamefont {Tschopp}\ and\ \citenamefont
  {Mcdowell}(2007)}]{Tschopp2007_1}%
  \BibitemOpen
  \bibfield  {author} {\bibinfo {author} {\bibfnamefont {M.~A.}\ \bibnamefont
  {Tschopp}}\ and\ \bibinfo {author} {\bibfnamefont {D.~L.}\ \bibnamefont
  {Mcdowell}},\ }\bibfield  {title} {\enquote {\bibinfo {title} {Asymmetric
  tilt grain boundary structure and energy in copper and aluminium},}\ }\href
  {\doibase 10.1080/14786430701455321} {\bibfield  {journal} {\bibinfo
  {journal} {Philosophical Magazine}\ }\textbf {\bibinfo {volume} {87}},\
  \bibinfo {pages} {3871--3892} (\bibinfo {year} {2007})}\BibitemShut {NoStop}%
\bibitem [{\citenamefont {Tschopp}\ and\ \citenamefont
  {McDowell}(2007)}]{Tschopp2007_2}%
  \BibitemOpen
  \bibfield  {author} {\bibinfo {author} {\bibfnamefont {M.~A.}\ \bibnamefont
  {Tschopp}}\ and\ \bibinfo {author} {\bibfnamefont {D.~L.}\ \bibnamefont
  {McDowell}},\ }\bibfield  {title} {\enquote {\bibinfo {title} {Structures and
  energies of $\sum3$ asymmetric tilt grain boundaries in copper and
  aluminium},}\ }\href {\doibase 10.1080/14786430701255895} {\bibfield
  {journal} {\bibinfo  {journal} {Philosophical Magazine}\ }\textbf {\bibinfo
  {volume} {87}},\ \bibinfo {pages} {3147--3173} (\bibinfo {year}
  {2007})}\BibitemShut {NoStop}%
\bibitem [{\citenamefont {Hadian}, \citenamefont {Grabowski},\ and\
  \citenamefont {Neugebauer}(2018)}]{Hadian2018}%
  \BibitemOpen
  \bibfield  {author} {\bibinfo {author} {\bibfnamefont {R.}~\bibnamefont
  {Hadian}}, \bibinfo {author} {\bibfnamefont {B.}~\bibnamefont {Grabowski}}, \
  and\ \bibinfo {author} {\bibfnamefont {J.}~\bibnamefont {Neugebauer}},\
  }\bibfield  {title} {\enquote {\bibinfo {title} {{GB}~code: A grain boundary
  generation code},}\ }\href {\doibase 10.21105/joss.00900} {\bibfield
  {journal} {\bibinfo  {journal} {Journal of Open Source Software}\ }\textbf
  {\bibinfo {volume} {3}},\ \bibinfo {pages} {900} (\bibinfo {year}
  {2018})}\BibitemShut {NoStop}%
\bibitem [{\citenamefont {Martyna}, \citenamefont {Tobias},\ and\ \citenamefont
  {Klein}(1994)}]{Martyna1994}%
  \BibitemOpen
  \bibfield  {author} {\bibinfo {author} {\bibfnamefont {G.~J.}\ \bibnamefont
  {Martyna}}, \bibinfo {author} {\bibfnamefont {D.~J.}\ \bibnamefont {Tobias}},
  \ and\ \bibinfo {author} {\bibfnamefont {M.~L.}\ \bibnamefont {Klein}},\
  }\bibfield  {title} {\enquote {\bibinfo {title} {Constant pressure molecular
  dynamics algorithms},}\ }\href {\doibase 10.1063/1.467468} {\bibfield
  {journal} {\bibinfo  {journal} {The Journal of Chemical Physics}\ }\textbf
  {\bibinfo {volume} {101}},\ \bibinfo {pages} {4177--4189} (\bibinfo {year}
  {1994})}\BibitemShut {NoStop}%
\bibitem [{\citenamefont {Hasnaoui}, \citenamefont {Derlet},\ and\
  \citenamefont {Swygenhoven}(2004)}]{Hasnaoui2004}%
  \BibitemOpen
  \bibfield  {author} {\bibinfo {author} {\bibfnamefont {A.}~\bibnamefont
  {Hasnaoui}}, \bibinfo {author} {\bibfnamefont {P.}~\bibnamefont {Derlet}}, \
  and\ \bibinfo {author} {\bibfnamefont {H.~V.}\ \bibnamefont {Swygenhoven}},\
  }\bibfield  {title} {\enquote {\bibinfo {title} {Interaction between
  dislocations and grain boundaries under an indenter {\textendash} a molecular
  dynamics simulation},}\ }\href {\doibase 10.1016/j.actamat.2004.01.018}
  {\bibfield  {journal} {\bibinfo  {journal} {Acta Materialia}\ }\textbf
  {\bibinfo {volume} {52}},\ \bibinfo {pages} {2251--2258} (\bibinfo {year}
  {2004})}\BibitemShut {NoStop}%
\bibitem [{\citenamefont {Feng}\ \emph {et~al.}(2015)\citenamefont {Feng},
  \citenamefont {Shang}, \citenamefont {Liu},\ and\ \citenamefont
  {Lu}}]{Feng2015}%
  \BibitemOpen
  \bibfield  {author} {\bibinfo {author} {\bibfnamefont {Y.}~\bibnamefont
  {Feng}}, \bibinfo {author} {\bibfnamefont {J.~X.}\ \bibnamefont {Shang}},
  \bibinfo {author} {\bibfnamefont {Z.~H.}\ \bibnamefont {Liu}}, \ and\
  \bibinfo {author} {\bibfnamefont {G.~H.}\ \bibnamefont {Lu}},\ }\bibfield
  {title} {\enquote {\bibinfo {title} {The energy and structure of (110) twist
  grain boundary in tungsten},}\ }\href {\doibase 10.1016/j.apsusc.2015.08.265}
  {\bibfield  {journal} {\bibinfo  {journal} {Applied Surface Science}\
  }\textbf {\bibinfo {volume} {357}},\ \bibinfo {pages} {262--267} (\bibinfo
  {year} {2015})}\BibitemShut {NoStop}%
\bibitem [{\citenamefont {Honeycutt}\ and\ \citenamefont
  {Andersen}(1987)}]{Honeycutt1987}%
  \BibitemOpen
  \bibfield  {author} {\bibinfo {author} {\bibfnamefont {J.~D.}\ \bibnamefont
  {Honeycutt}}\ and\ \bibinfo {author} {\bibfnamefont {H.~C.}\ \bibnamefont
  {Andersen}},\ }\bibfield  {title} {\enquote {\bibinfo {title} {Molecular
  dynamics study of melting and freezing of small lennard-jones clusters},}\
  }\href {\doibase 10.1021/j100303a014} {\bibfield  {journal} {\bibinfo
  {journal} {The Journal of Physical Chemistry}\ }\textbf {\bibinfo {volume}
  {91}},\ \bibinfo {pages} {4950--4963} (\bibinfo {year} {1987})}\BibitemShut
  {NoStop}%
\bibitem [{\citenamefont {Stukowski}\ and\ \citenamefont
  {Albe}(2010)}]{Stukowski2010}%
  \BibitemOpen
  \bibfield  {author} {\bibinfo {author} {\bibfnamefont {A.}~\bibnamefont
  {Stukowski}}\ and\ \bibinfo {author} {\bibfnamefont {K.}~\bibnamefont
  {Albe}},\ }\bibfield  {title} {\enquote {\bibinfo {title} {Extracting
  dislocations and non-dislocation crystal defects from atomistic simulation
  data},}\ }\href {\doibase 10.1088/0965-0393/18/8/085001} {\bibfield
  {journal} {\bibinfo  {journal} {Modelling and Simulation in Materials Science
  and Engineering}\ }\textbf {\bibinfo {volume} {18}},\ \bibinfo {pages}
  {085001} (\bibinfo {year} {2010})}\BibitemShut {NoStop}%
\bibitem [{\citenamefont {Stukowski}(2009)}]{Stukowski2009}%
  \BibitemOpen
  \bibfield  {author} {\bibinfo {author} {\bibfnamefont {A.}~\bibnamefont
  {Stukowski}},\ }\bibfield  {title} {\enquote {\bibinfo {title} {Visualization
  and analysis of atomistic simulation data with {OVITO}{\textendash}the open
  visualization tool},}\ }\href {\doibase 10.1088/0965-0393/18/1/015012}
  {\bibfield  {journal} {\bibinfo  {journal} {Modelling and Simulation in
  Materials Science and Engineering}\ }\textbf {\bibinfo {volume} {18}},\
  \bibinfo {pages} {015012} (\bibinfo {year} {2009})}\BibitemShut {NoStop}%
\bibitem [{\citenamefont {Rojas}, \citenamefont {Orhan},\ and\ \citenamefont
  {Ponga}(2021)}]{Rojas2021}%
  \BibitemOpen
  \bibfield  {author} {\bibinfo {author} {\bibfnamefont {D.~F.}\ \bibnamefont
  {Rojas}}, \bibinfo {author} {\bibfnamefont {O.~K.}\ \bibnamefont {Orhan}}, \
  and\ \bibinfo {author} {\bibfnamefont {M.}~\bibnamefont {Ponga}},\ }\bibfield
   {title} {\enquote {\bibinfo {title} {Dynamic recrystallization of silver
  nanocubes during high-velocity impacts},}\ }\href {\doibase
  10.1016/j.actamat.2021.116892} {\bibfield  {journal} {\bibinfo  {journal}
  {Acta Materialia}\ }\textbf {\bibinfo {volume} {212}},\ \bibinfo {pages}
  {116892} (\bibinfo {year} {2021})}\BibitemShut {NoStop}%
\bibitem [{\citenamefont {Galitskiy}, \citenamefont {Ivanov},\ and\
  \citenamefont {Dongare}(2018)}]{Galitskiy2018}%
  \BibitemOpen
  \bibfield  {author} {\bibinfo {author} {\bibfnamefont {S.}~\bibnamefont
  {Galitskiy}}, \bibinfo {author} {\bibfnamefont {D.~S.}\ \bibnamefont
  {Ivanov}}, \ and\ \bibinfo {author} {\bibfnamefont {A.~M.}\ \bibnamefont
  {Dongare}},\ }\bibfield  {title} {\enquote {\bibinfo {title} {Dynamic
  evolution of microstructure during laser shock loading and spall failure of
  single crystal al at the atomic scales},}\ }\href {\doibase
  10.1063/1.5051618} {\bibfield  {journal} {\bibinfo  {journal} {Journal of
  Applied Physics}\ }\textbf {\bibinfo {volume} {124}} (\bibinfo {year}
  {2018}),\ 10.1063/1.5051618}\BibitemShut {NoStop}%
\bibitem [{\citenamefont {Pang}\ \emph {et~al.}(2014)\citenamefont {Pang},
  \citenamefont {Zhang}, \citenamefont {Zhang}, \citenamefont {Xu},\ and\
  \citenamefont {Zhao}}]{Pang2014}%
  \BibitemOpen
  \bibfield  {author} {\bibinfo {author} {\bibfnamefont {W.~W.}\ \bibnamefont
  {Pang}}, \bibinfo {author} {\bibfnamefont {P.}~\bibnamefont {Zhang}},
  \bibinfo {author} {\bibfnamefont {G.~C.}\ \bibnamefont {Zhang}}, \bibinfo
  {author} {\bibfnamefont {A.~G.}\ \bibnamefont {Xu}}, \ and\ \bibinfo {author}
  {\bibfnamefont {X.~G.}\ \bibnamefont {Zhao}},\ }\bibfield  {title} {\enquote
  {\bibinfo {title} {Dislocation creation and void nucleation in {FCC} ductile
  metals under tensile loading: A general microscopic picture},}\ }\href
  {\doibase 10.1038/srep06981} {\bibfield  {journal} {\bibinfo  {journal}
  {Scientific Reports}\ }\textbf {\bibinfo {volume} {4}} (\bibinfo {year}
  {2014}),\ 10.1038/srep06981}\BibitemShut {NoStop}%
\bibitem [{\citenamefont {Shugaev}\ \emph {et~al.}(2021)\citenamefont
  {Shugaev}, \citenamefont {He}, \citenamefont {Levy}, \citenamefont {Mazzi},
  \citenamefont {Miotello}, \citenamefont {Bulgakova},\ and\ \citenamefont
  {Zhigilei}}]{Shugaev2021}%
  \BibitemOpen
  \bibfield  {author} {\bibinfo {author} {\bibfnamefont {M.~V.}\ \bibnamefont
  {Shugaev}}, \bibinfo {author} {\bibfnamefont {M.}~\bibnamefont {He}},
  \bibinfo {author} {\bibfnamefont {Y.}~\bibnamefont {Levy}}, \bibinfo {author}
  {\bibfnamefont {A.}~\bibnamefont {Mazzi}}, \bibinfo {author} {\bibfnamefont
  {A.}~\bibnamefont {Miotello}}, \bibinfo {author} {\bibfnamefont {N.~M.}\
  \bibnamefont {Bulgakova}}, \ and\ \bibinfo {author} {\bibfnamefont {L.~V.}\
  \bibnamefont {Zhigilei}},\ }\bibfield  {title} {\enquote {\bibinfo {title}
  {Laser-induced thermal processes: Heat transfer, generation of stresses,
  melting and solidification, vaporization, and phase explosion},}\ }in\ \href
  {\doibase 10.1007/978-3-030-63647-0_11} {\emph {\bibinfo {booktitle}
  {Handbook of Laser Micro- and Nano-Engineering}}}\ (\bibinfo  {publisher}
  {Springer International Publishing},\ \bibinfo {year} {2021})\ pp.\ \bibinfo
  {pages} {83--163}\BibitemShut {NoStop}%
\bibitem [{\citenamefont {Li}\ and\ \citenamefont {Guan}(2020)}]{Li2020}%
  \BibitemOpen
  \bibfield  {author} {\bibinfo {author} {\bibfnamefont {X.}~\bibnamefont
  {Li}}\ and\ \bibinfo {author} {\bibfnamefont {Y.}~\bibnamefont {Guan}},\
  }\bibfield  {title} {\enquote {\bibinfo {title} {Theoretical fundamentals of
  short pulse laser{\textendash}metal interaction: A review},}\ }\href
  {\doibase 10.1016/j.npe.2020.08.001} {\bibfield  {journal} {\bibinfo
  {journal} {Nanotechnology and Precision Engineering}\ }\textbf {\bibinfo
  {volume} {3}},\ \bibinfo {pages} {105--125} (\bibinfo {year}
  {2020})}\BibitemShut {NoStop}%
\bibitem [{\citenamefont {de~Ress{\'e}guier}\ \emph {et~al.}(2010)\citenamefont
  {de~Ress{\'e}guier}, \citenamefont {Loison}, \citenamefont {Lescoute},
  \citenamefont {Signor},\ and\ \citenamefont {Dragon}}]{de2010dynamic}%
  \BibitemOpen
  \bibfield  {author} {\bibinfo {author} {\bibfnamefont {T.}~\bibnamefont
  {de~Ress{\'e}guier}}, \bibinfo {author} {\bibfnamefont {D.}~\bibnamefont
  {Loison}}, \bibinfo {author} {\bibfnamefont {E.}~\bibnamefont {Lescoute}},
  \bibinfo {author} {\bibfnamefont {L.}~\bibnamefont {Signor}}, \ and\ \bibinfo
  {author} {\bibfnamefont {A.}~\bibnamefont {Dragon}},\ }\bibfield  {title}
  {\enquote {\bibinfo {title} {Dynamic fragmentation of laser shock-melted
  metals: some wxperimental advances},}\ }\href@noop {} {\bibfield  {journal}
  {\bibinfo  {journal} {Journal of Theoretical and Applied Mechanics}\ }\textbf
  {\bibinfo {volume} {48}},\ \bibinfo {pages} {957--972} (\bibinfo {year}
  {2010})}\BibitemShut {NoStop}%
\bibitem [{\citenamefont {Kumar}\ \emph {et~al.}(2003)\citenamefont {Kumar},
  \citenamefont {Suresh}, \citenamefont {Chisholm}, \citenamefont {Horton},\
  and\ \citenamefont {Wang}}]{Kumar2003}%
  \BibitemOpen
  \bibfield  {author} {\bibinfo {author} {\bibfnamefont {K.}~\bibnamefont
  {Kumar}}, \bibinfo {author} {\bibfnamefont {S.}~\bibnamefont {Suresh}},
  \bibinfo {author} {\bibfnamefont {M.}~\bibnamefont {Chisholm}}, \bibinfo
  {author} {\bibfnamefont {J.}~\bibnamefont {Horton}}, \ and\ \bibinfo {author}
  {\bibfnamefont {P.}~\bibnamefont {Wang}},\ }\bibfield  {title} {\enquote
  {\bibinfo {title} {Deformation of electrodeposited nanocrystalline nickel},}\
  }\href {\doibase 10.1016/s1359-6454(02)00421-4} {\bibfield  {journal}
  {\bibinfo  {journal} {Acta Materialia}\ }\textbf {\bibinfo {volume} {51}},\
  \bibinfo {pages} {387--405} (\bibinfo {year} {2003})}\BibitemShut {NoStop}%
\bibitem [{\citenamefont {Dongare}, \citenamefont {LaMattina},\ and\
  \citenamefont {Rajendran}(2011)}]{Dongare2011}%
  \BibitemOpen
  \bibfield  {author} {\bibinfo {author} {\bibfnamefont {A.~M.}\ \bibnamefont
  {Dongare}}, \bibinfo {author} {\bibfnamefont {B.}~\bibnamefont {LaMattina}},
  \ and\ \bibinfo {author} {\bibfnamefont {A.~M.}\ \bibnamefont {Rajendran}},\
  }\bibfield  {title} {\enquote {\bibinfo {title} {Atomic scale studies of
  spall behavior in single crystal cu},}\ }\href {\doibase
  10.1016/j.proeng.2011.04.598} {\bibfield  {journal} {\bibinfo  {journal}
  {Procedia Engineering}\ }\textbf {\bibinfo {volume} {10}},\ \bibinfo {pages}
  {3636--3641} (\bibinfo {year} {2011})}\BibitemShut {NoStop}%
\bibitem [{\citenamefont {Stukowski}, \citenamefont {Bulatov},\ and\
  \citenamefont {Arsenlis}(2012)}]{Stukowski2012}%
  \BibitemOpen
  \bibfield  {author} {\bibinfo {author} {\bibfnamefont {A.}~\bibnamefont
  {Stukowski}}, \bibinfo {author} {\bibfnamefont {V.~V.}\ \bibnamefont
  {Bulatov}}, \ and\ \bibinfo {author} {\bibfnamefont {A.}~\bibnamefont
  {Arsenlis}},\ }\bibfield  {title} {\enquote {\bibinfo {title} {Automated
  identification and indexing of dislocations in crystal interfaces},}\ }\href
  {\doibase 10.1088/0965-0393/20/8/085007} {\bibfield  {journal} {\bibinfo
  {journal} {Modelling and Simulation in Materials Science and Engineering}\
  }\textbf {\bibinfo {volume} {20}},\ \bibinfo {pages} {085007} (\bibinfo
  {year} {2012})}\BibitemShut {NoStop}%
\end{thebibliography}%
\end{document}